\documentclass[fleqn,10pt]{article}
\usepackage[utf8]{inputenc}
\usepackage[english]{babel}

\usepackage[scaled]{helvet}
\usepackage{amsmath,amsfonts,amssymb}

\DeclareMathOperator*{\argmin}{arg\,min}
\usepackage{graphicx,xcolor}
\usepackage{authblk}

\usepackage[left=1.35cm,
            right=1.35cm,
            top=2.25cm,
            bottom=2.25cm,
            headheight=12pt,
            a4paper]{geometry}
                
\usepackage[labelfont={bf,sf},
            labelsep=period,
            justification=justified]{caption}
\RequirePackage[colorlinks=true, allcolors=blue]{hyperref}

\usepackage[biblabel,nomove]{cite}
\newcommand{\keywords}[1]{\def\@keywords{#1}}

\usepackage[explicit]{titlesec}
\titleformat{\subsubsection}[runin]
  {\sffamily\bfseries\itshape}
  {\thesubsubsection}
  {0.5em}
  {#1}
\titleformat{\paragraph}[runin]
  {\normalfont\itshape}
  {\theparagraph}
  {0.5em}
  {#1}
\titlespacing*{\paragraph}{0pt}{0.5em}{0.5em}

\usepackage[T1]{fontenc}
\usepackage{wrapfig}
\usepackage{sidecap}
\sidecaptionvpos{figure}{t}
\usepackage{comment}

\usepackage[section]{placeins} % ADZ: force supp figs below "Supplementary Material" heading

\title{Visualizing the geometry of labeled high-dimensional data with spheres}
\author[1,*]{Andrew D Zaharia}
\author[1]{Anish S Potnis}
\author[2]{Alexander Walther}
\author[1,3]{Nikolaus Kriegeskorte}
\affil[1]{Mortimer B. Zuckerman Mind Brain Behavior Institute, Columbia University, New York, NY 10027}
\affil[2]{Realeyes, 2 Riding House St, London W1W 7FA, UK}
\affil[3]{Departments of Psychology, Neuroscience, and Electrical Engineering, Columbia University, New York, NY 10027}

\affil[*]{Address correspondence to: andrew.z@columbia.edu}


\graphicspath{{./figures/}}
\makeatletter
\def\ttl@useclass#1#2{%
  \@ifstar
    {\ttl@labelfalse\@dblarg{#1{#2}}}% {\ttl@labelfalse#1{#2}[]}%
    {\ttl@labeltrue\@dblarg{#1{#2}}}}
  \renewenvironment{abstract}{
    \begin{center}
      {\bfseries \Large\abstractname\vspace{\z@}}
    \end{center}
  }
\makeatother

\newcommand{\beginsupplement}{%
        \setcounter{table}{0}
        \renewcommand{\thetable}{S\arabic{table}}%
        \setcounter{figure}{0}
        \renewcommand{\thefigure}{S\arabic{figure}}
     }
\newcommand*\autorefsupp[1]{Supplementary \autoref{#1}}

\usepackage{array}
\newcolumntype{*}{>{\global\let\currentrowstyle\relax}}
\newcolumntype{^}{>{\currentrowstyle}}
\newcommand{\rowstyle}[1]{\gdef\currentrowstyle{#1}%
  #1\ignorespaces
}

\usepackage{xcolor}
\usepackage{soul}
\setul{0pt}{.4pt}
\setul{-.6ex}{.3ex}
\newcommand{\revadd}[1]{#1} % disabled
\begin{document}
\flushbottom
\maketitle
\thispagestyle{empty}

% "Articles begin with an unreferenced abstract (typically 150 words)..."
\begin{abstract} % RIGHT NOW: 164 WORDS
\normalsize\bfseries
\noindent Data visualizations \revadd{can} summarize high-dimensional distributions in two or three dimensions. Dimensionality reduction entails a loss of information, and what is preserved differs between methods. Existing methods preserve the local or the global geometry of the points, and most techniques do not consider labels. Here we introduce  ``\mbox{hypersphere2sphere}'' (H2S), a new method that aims to visualize not the points, but the relationships between the labeled distributions. H2S fits a hypersphere to each labeled set of points \revadd{in the high-dimensional space of the original data} and visualizes each hypersphere as a sphere in 3D (or a circle in 2D). H2S perfectly captures the geometry of up to 4 \revadd{high-dimensional} hyperspheres \revadd{using spheres} in 3D (or 3 \revadd{using circles} in 2D), and approximates the geometry for larger numbers of distributions, matching the sizes (radii), and the pairwise separations (between-center distances) and \revadd{the} overlaps (along the center-connection line). The resulting visualizations are robust to sampling imbalances. Leveraging labels and the sphere as the simplest geometrical primitive, H2S provides an important addition to the toolbox of visualization techniques.
\end{abstract}

%%%%%%%%%%%%%%%%%%%%%%%%%%%%%%%%%%%%%% INTRODUCTION %%%%%%%%%%%%%%%%%%%%%%%%%%%%%%%%%%%%%%
% 671 words (not that it matters too much). (Deep-Z is 729, SIFT is 897)
% Introduction and Discussion are brief and focused...

\section*{Introduction}
% dimensionality reduction techniques make tradeoffs
% compare existing dimensionality reduction techniques. they emphasize different pieces of information

\begin{figure}[b]
  \centering
  \includegraphics[width=180mm]{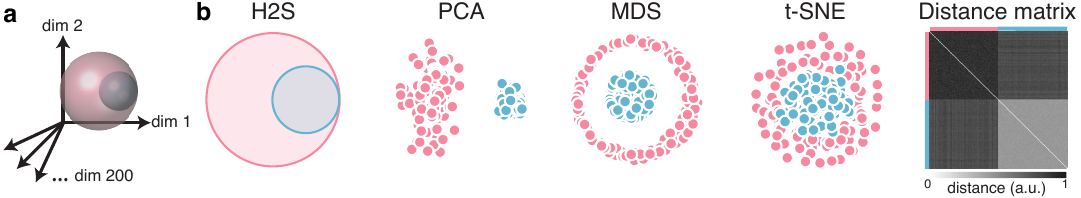}
  \caption{\label{fig:problem}\textbf{How should labeled high-dimensional distributions be visualized?} Visualization of 200-dimensional simulated data with the proposed method hypersphere-to-sphere (H2S) and traditional methods (PCA, MDS, \textit{t}-SNE). (\textbf{a}) The ground-truth data-generating distributions here are uniform balls in 200 dimensions. The larger 200D ball contains the smaller one entirely. Their bounding hyperspheres touch in a single point. The data set consists of 100 points sampled independently from each 200D-ball. (\textbf{b}) Different visualization methods. Hypersphere2sphere (H2S, proposed here) visualizes the 200D balls as circles, revealing the enclosure relationship, the radii of the balls, and the offset separating their centers. Principal component analysis (PCA), multidimensional scaling using metric stress (MDS), and \textit{t}-distributed stochastic neighbor embedding (\textit{t}-SNE) do not reveal the global geometry in this example. The matrix of pairwise point distances \revadd{describes} the geometry completely, but does not convey it intuitively.}
\end{figure}

Data visualization engages our visual system's capacity for parallel processing, enabling us to understand a complex set of relationships at a glance. Visualization of high-dimensional data typically involves embedding the data in a 2- or 3-dimensional space, such that every high-dimensional data point is displayed in the visualization. The embedding, however, requires dimensionality reduction, which inevitably entails a loss of information. \revadd{The resulting visualizations are, therefore, simplified representations of the original high-dimensional data distributions.} Different methods make different \revadd{choices} in terms of what information to preserve and what to discard. The most commonly used visualization methods reduce dimensionality without making use of data labels.

% \begin{SCfigure}
\begin{wrapfigure}{r}{0.5\textwidth}%\vspace{-22mm}
% \begin{figure}[ht]
  \centering
  \includegraphics[width=88mm]{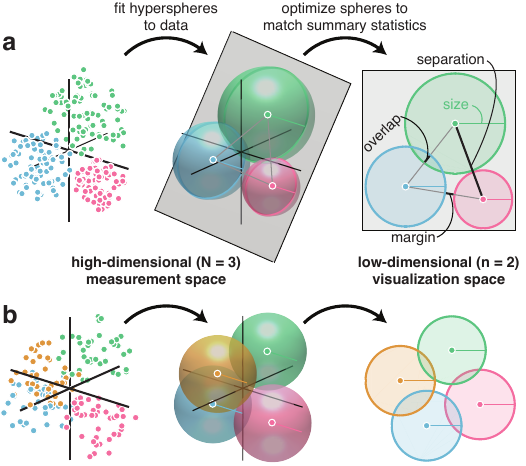}
  \caption{\textbf{H2S fits hyperspheres enclosing high-dimensional labeled distributions and visualizes them as spheres.}
  In two example cases (\textbf{a}, \textbf{b}), the true distributions (not shown) are uniform within hyperspheres in $N = 3$ dimensions (three labeled distributions in \textbf{a}, four in \textbf{b}). The data (left column) are drawn i.i.d. from these distributions. \revadd{The enclosing} hyperspheres (middle column) in the original space (3D here) \revadd{represent the location and extent of the distributions, not their density.} The visualization (right column) consists of a circle in $n = 2$ dimensions for each distribution (\textbf{a}, right column). \revadd{In this case,} the H2S visualization is \revadd{equivalent to} a planar slice defined by the three centers of the spheres. Each hypersphere is represented by a circle marking the intersection of the hypersphere and the visualization plane. The resulting 2D set of circles \textit{perfectly} represents the pairwise distances between the hypersphere centers and the hypersphere radii and overlaps. In general, H2S can perfectly represent $n+1$ \revadd{enclosing hyperspheres} in an $n$-dimensional visualization space (4 in a 3D or 3 in a 2D visualization), since an $n$-dimensional hyperplane can perfectly capture the radii and center separations for all $n+1$ hyperspheres. (\textbf{b}) \revadd{In order to generalize to larger numbers of classes, H2S optimizes the ensemble of visualization spheres such that their summary statistics most closely match the hypersphere summary statistics}: sphere radii match hypersphere radii, sphere center separations match hypersphere center separations, and sphere overlaps match hypersphere overlaps (defined for each pair of classes as the length along the line connecting the centers).}\label{fig:intro}
% \end{figure}
\end{wrapfigure}
% \end{SCfigure}

Several popular dimensionality reduction techniques attempt to preserve the geometry of the data (\autoref{fig:problem}a). Principal Component Analysis (PCA) preserves the geometry of the high-dimensional data exactly, but only for the two or three dimensions explaining the most variance (\autoref{fig:problem}b). Multidimensional Scaling (MDS\cite{Young1938,Torgerson1952,Shepard1962}) finds an embedding of the points in the visualization space that best matches the high-dimensional distances (\autoref{fig:problem}b).

Other techniques preserve local neighborhood relationships rather than global geometry. Sammon Mapping\cite{Sammon1969} is a variant of MDS that prioritizes getting local relationships right. Isomap\cite{Tenenbaum2000a} and Locally Linear Embedding\cite{Roweis2000} preserve distances on a manifold defined by local neighborhood relationships in the data. The \textit{t}-distributed stochastic neighbor embedding (\textit{t}-SNE \cite{VanderMaaten2008}, \autoref{fig:problem}b) and the Uniform Manifold Approximation and Projection (UMAP, \cite{McInnes2018a}) attempt to find transformations that preserve stochastically defined neighborhood relationships. Despite their frequent application to labeled data, each of these techniques embeds the individual data points in visualization space without using labels.

\revadd{Often,} we have labels for the data we wish to visualize and are interested in the relationships among the labeled distributions. We can render visualized points with colors signifying their class labels. Using a label-blind (unsupervised) embedding and coloring by class label has the advantage that the embedding is not biased by the labels. Label-aware (supervised) analyses can reveal complementary information.\footnote{\revadd{See https://umap-learn.readthedocs.io/en/latest/supervised.html for a label-aware, supervised extension of UMAP.}} Here we introduce a new label-aware visualization method with an alternative aim: \revadd{to represent each class as a \textit{single visual object}, directly conveying the relationships among the labeled distributions rather than those among individual points}.

We introduce \mbox{``hypersphere2sphere''} (H2S\footnote{Available for MATLAB, Python version forthcoming.}, \autoref{fig:intro}), \revadd{a visualization method }for labeled high-dimensional data that does not embed individual data points in the visualization space, but instead represents each distribution as a \revadd{sphere} in the visualization space. \revadd{Each labeled distribution is first captured by an enclosing hypersphere in the original high-dimensional data space. These hyperspheres are not probability density models, but meant to capture the locations and extents of the distributions, enclosing almost all of their probability mass.} The set of hyperspheres and their relationships are then conveyed by a corresponding ensemble of spheres in 3D (or circles in 2D).

A hypersphere is the bounding surface of a ball in $N$ dimensions. \revadd{A ball is a set of all points within a radius $r$ of a central point. An $N$-dimensional ball constitutes an extremely simple and generic model of the location and extent of a distribution. Each hypersphere is estimated so as to enclose almost all of the distribution for a label (\autoref{fig:intro}a). The enclosing hypersphere ignores the shape of the distribution, but conveys its location and extent.} The radii, pairwise distances between centers, and pairwise overlaps of the hyperspheres are then expressed by the corresponding parameters of the spheres in the 3D embedding (or the circles in a 2D one, \autoref{fig:intro}b). We also visualize the values of the summary statistics (radii, inter-center distances, and overlaps) using a variant of the Hinton diagram \cite{Hinton1991}, alongside statistical comparisons among them.

H2S is useful for representing labeled, high-dimensional data, as are common in many fields of science. \revadd{The underlying distributions need not be ball-uniform, isotropic, or bounded for the H2S visualization to be useful in practice. An example of an unbounded distribution is an isotropic Gaussian distribution. In a high-dimensional space,} isotropic Gaussian distributions and their overlap relationships are well characterized by enclosing hyperspheres. \revadd{Although an isotropic Gaussian has unbounded support, extending infinitely in all directions, almost all of the probability mass is confined to a ball. The enclosing hyperspheres can be defined to contain 95\% or 99\% of the probability mass and provide an excellent representation of the relative locations, extents, and overlaps of the distributions. Not only does such a hypersphere enclose most of the mass, but most of the mass is within a small distance inward from the surface. }If the distributions are not isotropic, the enclosing hyperspheres still provide an informative summary of the relative locations and extents, \revadd{with the caveat that} overlap\revadd{ping spheres in the H2S visualization then do not imply overlapping high-dimensional distributions: An enclosing hypersphere is guaranteed to capture most of the probability mass of the distribution it represents, but the distribution is not guaranteed to extend throughout the enclosing hypersphere; it might have density zero in a portion of the hypersphere. Nonoverlapping visualization spheres do indicate nonoverlapping high-dimensional distributions.}

To demonstrate and validate H2S, we applied it to simulated data, human brain-activity patterns elicited by visual images, and internal representations in a neural network model. For comparison, we present \revadd{visualizations of the same data using} PCA, MDS, and \textit{t}-SNE. H2S provides a useful complement to these methods for conveying the relationships among the enclosing hyperspheres of the labeled distributions. In contrast to alternative methods, H2S is robust to imbalances in the numbers of samples between classes. It is well suited to provide a global characterization of class relationships even when the number of data points is much smaller than the dimensionality of the data.

%%%%%%%%%%%%%%%%%%%%%%%%%%%%%%%%%%%%%%   RESULTS   %%%%%%%%%%%%%%%%%%%%%%%%%%%%%%%%%%%%%%
% 271+360+251+239+242+404+292+206 = 2265 words
% the Results section usually contains a general description of the method followed by its validation, and the online Methods section provides all technical details necessary for the independent reproduction of the methodology, without referring to a chain of bibliographical references.
% The Results and online Methods should be divided by topical subheadings; the Discussion may contain subheadings at the editors' discretion. As a guideline, Articles allow up to 40 references.
\section*{Results}
\revadd{A hypersphere provides a simple characterization of the location and extent of a high-dimensional distribution, requiring only }%271 words (including subheading)
a radius and $N$ center coordinates, where $N$ is the dimensionality of the original data space. The visual language of spheres in 3D can efficiently and intuitively convey the summary statistics \revadd{that define the geometrical relationships among the hyperspheres: their sizes, separations, and overlaps. The size of each hypersphere is defined as its radius. For each pair of hyperspheres, the separation is defined as the between-center distance and the overlap is defined as the length of the segment of the center-connection line contained in both hyperspheres.} For a 3D visualization, spheres can perfectly express the summary statistics of up to four hyperspheres (\autoref{fig:intro}a), \revadd{independent of the original dimensionality $N$}. Likewise for a 2D visualization, circles can perfectly express the summary statistics of up to three hyperspheres. For larger numbers of distributions, the summary statistics are expressed approximately.

\revadd{Using hyperspheres to represent high-dimensional distributions is well motivated when any of the following three conditions hold: (1) The goal is to give a first sense of relative locations and extents of the distributions. (2) The number of data points does not support estimation of detailed density models given the dimensionality of the space. (3) The distributions are approximately isotropic. Condition (3) will seldom hold, but (1) and/or (2) provide a motivation for H2S in many common scenarios.}

\revadd{If we have ample data ($P>>N$), we may prefer a more detailed characterization of the distributions, capturing the potentially nonisotropic shapes of their density functions. However, even a nonisotropic Gaussian model in $N$ dimensions has $N+(N^2+N)/2$ parameters and so may be difficult to estimate. Even if we have enough data to estimate a complex model in $N$ dimensions, visualizing the complex shape of the distribution is often not possible and it is still useful to characterize the relative locations and expanses of the distributions with enclosing hyperspheres. Moreover, many nonuniform densities converge to a hypersphere in high dimensions. As an example of a nonuniform density with unbounded support, consider an isotropic Gaussian density. In a low-dimensional space, the high probability density at the center and the long tail of the Gaussian represent marked differences from the uniform $N$-ball distribution. In a high-dimensional space, however, $N$-ball and Gaussian distributions converge in the sense that most points fall within a small margin of a hypersphere (\autorefsupp{supp:histograms}). Other density models also converge to a hypersphere in high dimensions. This is true, for example, for a uniform hypercubic distribution, despite its anisotropy. More importantly for practical applications, these distributions are well approximated by a hypersphere in $N$ dimensions when $N$ is high.}

\revadd{If we do not have ample data ($P$ on the order of $N$), estimating complex $N$-dimensional density models is unrealistic. In many fields, we frequently have even fewer observed points than dimensions. In neuroscience, for example, we may study the responses of $N$ neurons to $P$ stimuli, where $P<N$. In this frequent scenario, it is not possible to estimate more complex density models without strong assumptions. A hypersphere can still capture where and how spread out the data points are. Despite its simplicity, even a hypersphere model is arguably too flexible in the $P<N$ scenario. For $P$ points in $N$ dimensions, all points lie on a hypersphere when $P=N+1$. When $P\leq N$, then there are infinitely many hyperspheres, each including all points. A hypersphere including the $P$ points means that all points lie on the surface of the ball, not inside the ball. This motivates finding the minimal enclosing hypersphere, the smallest hypersphere that contains all the points. More generally, we are interested in estimating an approximate minimal enclosing hypersphere that encloses almost all of the underlying distribution's probability mass.} 

\subsubsection*{Hypersphere parameter estimation.} %360 words (including subheading)
\label{ssec:parest}
The first step of H2S is to estimate the center $\hat{\bf c}_i$ and radius $\hat{r}_i$ of the hypersphere that best describes the $i$th labeled distribution in the original high-dimensional space. A number of estimators have been proposed for the minimum enclosing ball of a set of points \cite{Kumar2003,Badoiu2003,Fischer2003,Ding2019}. Alternative estimators for the center and radius differ in their assumptions and in their statistical and computational efficiency (\autorefsupp{tab:estimators}). An estimator is suitable if it is within our computational budget and its assumptions hold approximately for the data.

We implemented and validated a range of estimators. Assuming an $N$-ball uniform distribution model, the maximum likelihood (ML) estimate is the smallest $N$-ball that encloses all points. The smallest $N$-ball has the highest density, thus maximizing the likelihood. The parameters of the smallest enclosing $N$-ball can be efficiently approximated by optimization \cite{Ritter1990,Larsson2008}. Because the ML estimate here is the smallest hypersphere possible, given the data, it will underestimate the radius of the true distribution for finite samples. The negative bias of the ML estimate of the radius can be severe when the number of samples is small relative to the number of dimensions. To infer the joint posterior over the center and radius, we implemented a Markov-Chain Monte Carlo (MCMC) estimator. This estimator \revadd{has attractive properties}, but requires much computation.

\revadd{The fact that many distributions converge to} Gaussian distributions in high dimensions motivates estimation of the center as we would for a Gaussian, where the ML estimate is simply the mean of all points. \revadd{We can then estimate the radius as the median (50th percentile) or the 95th percentile of the empirical distance-to-center distribution (using the mean as the center estimate). Choosing the 95th percentile ensures that the hypersphere encloses 95\% of the points in our data set. In high-dimensional scenarios, however, the radius estimates when using the median or the 95th percentile of the distance-to-center distribution will be almost identical. The reason for this insensitivity of the chosen percentile is that the distance-to-center has a small variance relative to the radius (\autoref{supp:histograms}). The small variance results from the law of large numbers (of dimensions here, across which the squared deviations to the center are summed). This radius estimator based on the empirical distance-to-center distribution combines speed and accuracy across a range of simulated distributions, dimensionalities, and numbers of samples (\autorefsupp{supp:performance}). See \nameref{supp:methods} and, in particular, \autoref{eqn:deviation} for details. If only the distances among data points are provided, instead of the data points themselves, our implementation of H2S uses distance-matrix-based radius and center-separations estimators (\autoref{eqn:rdist} in \nameref{supp:methods}).}

\subsubsection*{Hypersphere summary statistics.} %251 words (not including equation lines)
The hypersphere summary statistics measure the size of each hypersphere and the separation and overlap for each pair of hyperspheres. The size of the $i$th hypersphere is defined as its radius $r_i$. The separation between the $i$th and $j$th hyperspheres is defined as the distance $d_{ij}$ between their centers. The distances are estimated as follows:
\begin{align}
\label{eqn:dists}
  \hat{d}_{ij} &= || \hat{\bf c}_i - \hat{\bf c}_j ||_2,
\end{align}
where $\hat{\bf c}_i$ and $\hat{\bf c}_j$ are estimates of the centers ${\bf c}_i$ and ${\bf c}_j$: the \revadd{empirical means of the samples for the two distributions (except when using MCMC or ML estimators, which jointly estimate the center and radius parameters, see \nameref{supp:methods})}.
The margin $m_{ij}$ between two hyperspheres is defined as the distance between the centers minus the radii. It is estimated from the distance and radius estimates:
\begin{align}
\label{eqn:margins}
  \hat{m}_{ij} &= \hat{d}_{ij} - \hat{r}_i - \hat{r}_j
\end{align}
The overlap between two hyperspheres is defined as the negative margin: $-m_{ij}$. Note that the sizes, separations, and overlaps are one-dimensional quantities that have units of length in the original space. Size and overlap could alternatively be measured by volumes. However, units of volume depend on dimensionality and so cannot be directly compared between the data space and the visualization space. Measuring lengths in the original space and conveying them with lengths in the visualization space is highly interpretable.

%However, the effect of dimensionality has to be kept in mind in interpretation. For example, if an $N$-ball has twice the radius of another, its volume will be $2^N$ times larger. Using lengths is robust to the number of dimensions of the original space, which often reflects accidental properties of a data set, such as limitations on the number of channels that could be measured. A larger random sample of channels would enable more precise estimates of the radii, separations, and overlaps. Because summary statistic ratios are invariant to dimensionality changes, the resulting visualizations are robust.

\subsubsection*{Low-dimensional sphere embedding.} %239 words (not including equation lines)
\label{ssec:embed}
We jointly optimize the low-dimensional visualized sphere (or circle) parameters to minimize the deviation between the summary statistics of the hyperspheres (data) and the summary statistics of the spheres (visualization). Specifically, the optimization procedure finds the low-dimensional center positions $\tilde{\bf c}_1, \ldots, \tilde{\bf c}_T$ and radii $\tilde{r}_1, \ldots, \tilde{r}_T$ that minimize the error $E$, where $T$ is the number of classes. The error $E$ is a function of all the distances $\tilde{d}_{ij}$, margins $\tilde{m}_{ij}$, and radii $\tilde{r}_i$ of the visualization spheres and the corresponding summary statistics estimated from the high-dimensional data ($\hat{d}_{ij}$, $\hat{m}_{ij}$, and $\hat{r}_i$):
\begin{align}
\label{eqn:simpleobj}
  E(\tilde{\bf c}_1, \ldots, \tilde{\bf c}_T, \tilde{r}_1, \ldots, \tilde{r}_T) &= \sum_{i=1}^{T-1} \sum_{j=i+1}^T ( \tilde{d}_{ij} - \hat{d}_{ij} )^2 + 
       \sum_{i=1}^{T-1} \sum_{j=i+1}^T \left( \tilde{m}_{ij} - \hat{m}_{ij} \right)^2 + 
       \sum_{i=1}^T                    \left( \tilde{r}_i    - \hat{r}_i    \right)^2
\end{align}

\begin{figure}[b!]
  \centering
  \includegraphics[width=180mm]{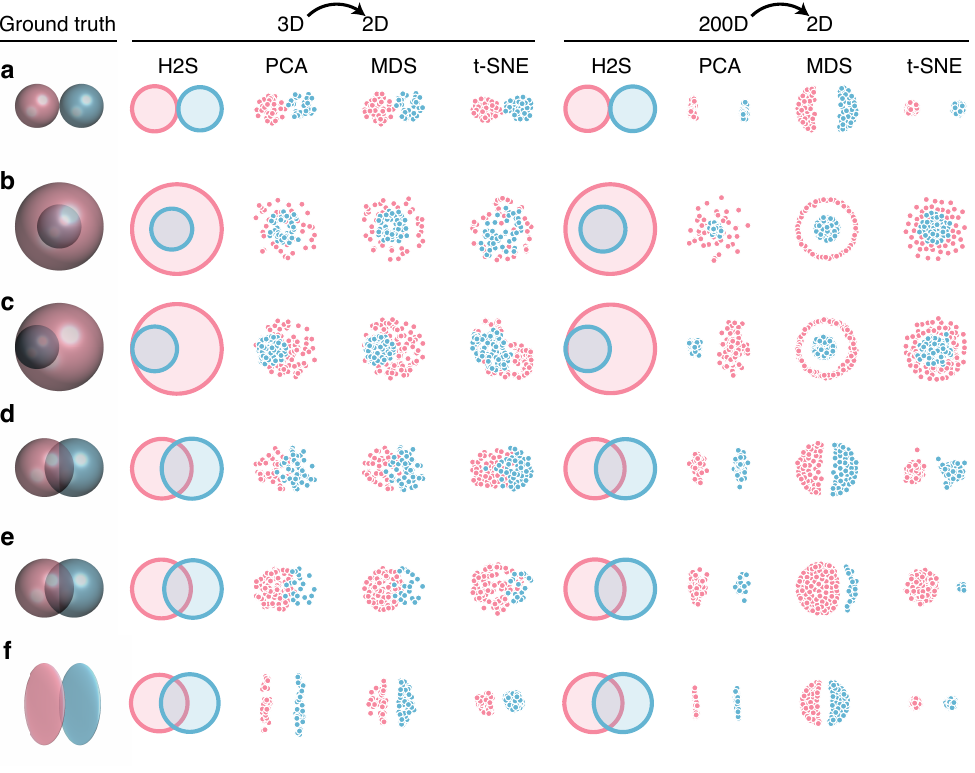}
  \caption{\label{fig:examples}\textbf{H2S is a more consistent visualization than PCA, MDS, and \textit{t}-SNE on 3D and 200D synthetic examples.} The leftmost column shows 3D renderings of uniform 3-ball distributions generating the points to the right. The next four columns from the left show 2D embeddings of the 3D distributions, and the four rightmost columns show 2D embeddings of equivalent 200-dimensional distributions. (\textbf{a}) Two touching equal-radius balls. (\textbf{b}) Two concentric balls with one having half the radius of the other. (\textbf{c}) One larger ball enclosing a smaller one, touching at one surface point. (\textbf{d}, \textbf{e}) Two intersecting balls with equal radii and the same number of points (\textbf{d}), or a different number of points (\textbf{e}, 100 red, 20 blue). \revadd{(\textbf{f}) Two uniform ellipsoid "pancake" distributions, each having their minor axes 8 times smaller than their major two (unit) axes.}}
\end{figure}
% \TODO{add hypercubes example}

\noindent
\revadd{Note that the margins $m_{ij}$ are a function of the center separations $d_{ij}$ and radii $r_i$, and that the same relationship holds for their estimates $\hat{m}_{ij}$, $\hat{d}_{ij}$, $\hat{r}_i$, and for the corresponding visualization parameters $\tilde{m}_{ij}$, $\tilde{d}_{ij}$, $\tilde{r}_i$ (Eq. \ref{eqn:margins}). A simplified error function would omit the margin-related middle term of the right-hand side of Eq. \ref{eqn:simpleobj}. Both of these error functions are $0$ when the estimated center separations and radii are perfectly expressed in the visualization. However, the total squared error will differ between the two error functions when only an approximate solution can be found. The motivation for including the margin-related middle term is that sizes, separations, and margins are perceptually separate properties of the visualization, all of which we would like to represent well. (See Eq. \ref{eqn:generalizedobj} for a family that generalizes these two choices of error function.)}

The error $E$ is not convex with respect to the visualization centers and radii. We minimize it by gradient descent to find a local minimum. First, we initialize the radii of the visualization spheres to be the same as the estimated high-dimensional radii and use MDS with metric stress \cite{Young1938,Torgerson1952} to arrange the centers. When the number of classes $T$ is at most $n+1$ (at most one greater than the visualization dimension $n \in \{2,3\}$), the distances between centers are perfectly expressed. The radii and margins are then also perfectly expressed, and the error vanishes: $E=0$. When $T$ is greater than $n+1$, we use gradient descent to find a local minimum of $E$, using a quasi-Newton (BFGS) algorithm (MATLAB's \textit{fminunc}) to optimize the radii and low-dimensional center positions (see \nameref{supp:methods}).

\revadd{To convey visualization error, each panel displays a bar at the bottom right indicating the maximum absolute error across all summary statistics, providing an at-a-glance upper bound on visualization fidelity.}

\subsubsection*{Validation of H2S using synthetic data.} %242 words

% \begin{SCfigure}
\begin{wrapfigure}{r}{0.5\textwidth}
% \begin{figure}[ht]
\vspace{-5mm}
  \centering
  \includegraphics[width=88mm]{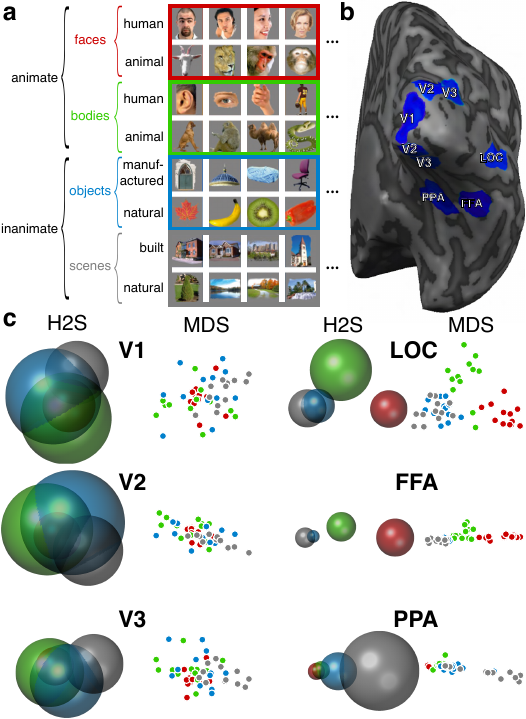}
  \caption{\textbf{Brain activity pattern distributions associated with visual object categories are confined to nonoverlapping hyperspheres at later stages of visual processing.}\\
  (\textbf{a}) Visual stimuli of different categories were presented to human subjects during a functional brain imaging experiment\cite{Walther2015}. (\textbf{b}) Inflated cortical surface reconstruction of the right hemisphere of one subject with six visual regions of interest (blue) whose representations of the images were separately analyzed. For each region, a 320-dimensional response pattern was estimated for each image. (\textbf{c}) H2S (left) and MDS (right) representations of faces (red), bodies (green), objects (blue), and scenes (gray), for the six visual areas in (\textbf{b}). At early stages (V1-3) the response patterns associated with different categories are entangled \cite{DiCarlo2007}, falling into overlapping enclosing hyperspheres. \revadd{As the visual information is re-represented at later stages of the visual processing hierarchy (lateral occipital complex, LOC; fusiform face area, FFA; parahippocampal place area, PPA), each category gains its own territory in response pattern space. The overlap of the enclosing hyperspheres is progressively reduced and in some cases (e.g. faces in LOC and FFA) a significant margin between the hyperspheres appears}.\label{fig:trans62}}
% \end{figure}
\end{wrapfigure}
% \end{SCfigure}

To demonstrate the fundamental properties of H2S, we used it to create 2D visualizations of simulated data in 3D and 200D where ground truth is known. We compared H2S to other visualization techniques: PCA, MDS, and \textit{t}-SNE. For equal radius $N$-balls touching at one point (\autoref{fig:examples}a), all methods give similar results when $N=3$ (left). When $N=200$ (right), however, all but H2S exaggerate the spatial separation between the two distributions, and PCA and MDS \revadd{suggest anisotropic distributions}. For concentric balls with different radii (\autoref{fig:examples}b), all visualizations reveal \revadd{that} the smaller (blue) distribution \revadd{is} surrounded by the larger (red) one. However, the configuration of the distributions changes substantially as the dimensionality grows for all methods except H2S, which accurately represents the geometry regardless of dimensionality.

\begin{figure}[b!]
  \centering
  \includegraphics[width=180mm]{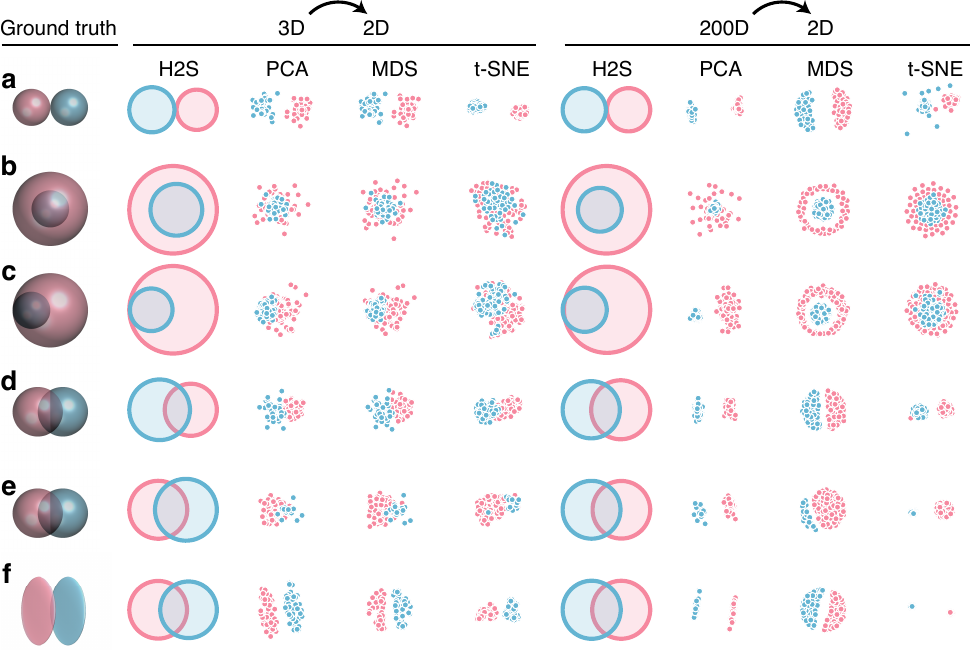}
  \caption{\label{supp:examples-gaussian}\revadd{\textbf{Consistency of visualizations for 3D and 200D Gaussian examples.} This figure replicates \autoref{fig:examples}, replacing the generating distributions, which were uniform balls in the former, with Gaussians in this one. The leftmost column shows 3D renderings of 3D Gaussian distributions generating the points to the right. The next four columns from the left show 2D embeddings of the 3D distributions, and the four rightmost columns show 2D embeddings of equivalent 200-dimensional Gaussian distributions. (\textbf{a}) Two (95\% level-set) "touching" equal-sigma Gaussians. (\textbf{b}) Two concentric Gaussians with one having half the sigma of the other. (\textbf{c}) One larger Gaussian enclosing a smaller one, "touching" at one surface point on their 95\% level-set. (\textbf{d}, \textbf{e}) Two intersecting Gaussians with equal sigma and the same number of points (\textbf{d}), or a different number of points (\textbf{e}, 100 red, 20 blue). (\textbf{f}) Two elongated ellipsoid "pancake" Gaussian distributions, each having their minor axes 8 times smaller than their major two (unit) axes.}}
\end{figure}

When the smaller inner ball is touching the larger ball at one point (\autoref{fig:examples}c), H2S reflects this in both the 3D and 200D cases, whereas the other visualizations do not provide an accurate picture of the global geometry of the two hyperspheres in 200D. Similarly, when two equal-radius hyperspheres intersect (\autoref{fig:examples}d), the overlap is correctly reflected, regardless of dimensionality in H2S, but the other visualizations suggest no overlap in 200D. The embedding geometry is strongly dependent on the relative number of samples for each class in MDS and \textit{t}-SNE. H2S, by contrast, is robust to the imbalance of the number of samples (\autoref{fig:examples}e, 100 red points, 20 blue). As expected, when the underlying distributions are anisotropic (\autoref{fig:examples}f), PCA and MDS capture the anisotropy in low and high dimensions, while t-SNE and H2S remain invariant to it. Finally, when creating Gaussian distribution variants of these synthetic data tests, we find the visualizations among all the methods to be largely similar (\autorefsupp{supp:examples-gaussian}).

\subsubsection*{Cognitive neuroscience application.} %404 words
\label{ssec:trans62}

We applied H2S to a neuroscientific dataset. Twenty-four human subjects viewed 48 images of objects belonging to four different categories (faces, bodies, objects, and places, \autoref{fig:trans62}a) while their brain activity was measured with functional magnetic resonance imaging (fMRI)\cite{Walther2015}. For each image, a detailed brain response pattern was estimated. The patterns were composed of 2mm-wide cubic measurement regions, called voxels, each reflecting the blood-oxygen-level-dependent response, which is correlated with local neural population activity. Analyses focused on six visual areas, defined as 320-voxel regions of interest (\autoref{fig:trans62}b).

Given the large number of responses for each region (320) and the much smaller number of images (48), estimating the complex high-dimensional category-defined distributions of activity patterns is unrealistic. However, H2S can reveal the expanse of the distribution for each category in the 320-dimensional space of response patterns, as well as the relationships among the different distributions. This can reveal how distinctly the brain represents different images, within and between categories, at different stages of visual processing.

\revadd{The human brain is thought to perform object recognition by progressively transforming representations such that the distributions of response patterns corresponding to categories become separable to downstream cortical areas that perform decoding using linear \cite{mcculloch43a} or radial-basis function classifiers \cite{Poggio1989}. H2S here visualizes the progressive disentangling \cite{DiCarlo2007}.} In early visual areas (V1-V3), including primary visual cortex (V1), different categories of image are ``entangled'' in the same territory of the multivariate response space. Although each image may elicit a unique response pattern, implying that the category distributions do not locally overlap, their global entanglement precludes a simple readout. 

The enclosing hyperspheres overlap significantly in early visual areas (all overlaps in V1 and V2 are significant: $P < 0.05$ for scenes vs faces and scenes vs bodies, $P < 0.001$ for all other pairs of categories). In V3, the categories are somewhat less entangled and the overlap for scenes vs faces and scenes vs bodies are not significant (other overlaps in V3 are still significant $P < 0.001$, see \autoref{fig:trans62}c and \autorefsupp{supp:trans62}).

\begin{figure}[b!]
  \centering
  \includegraphics[width=180mm]{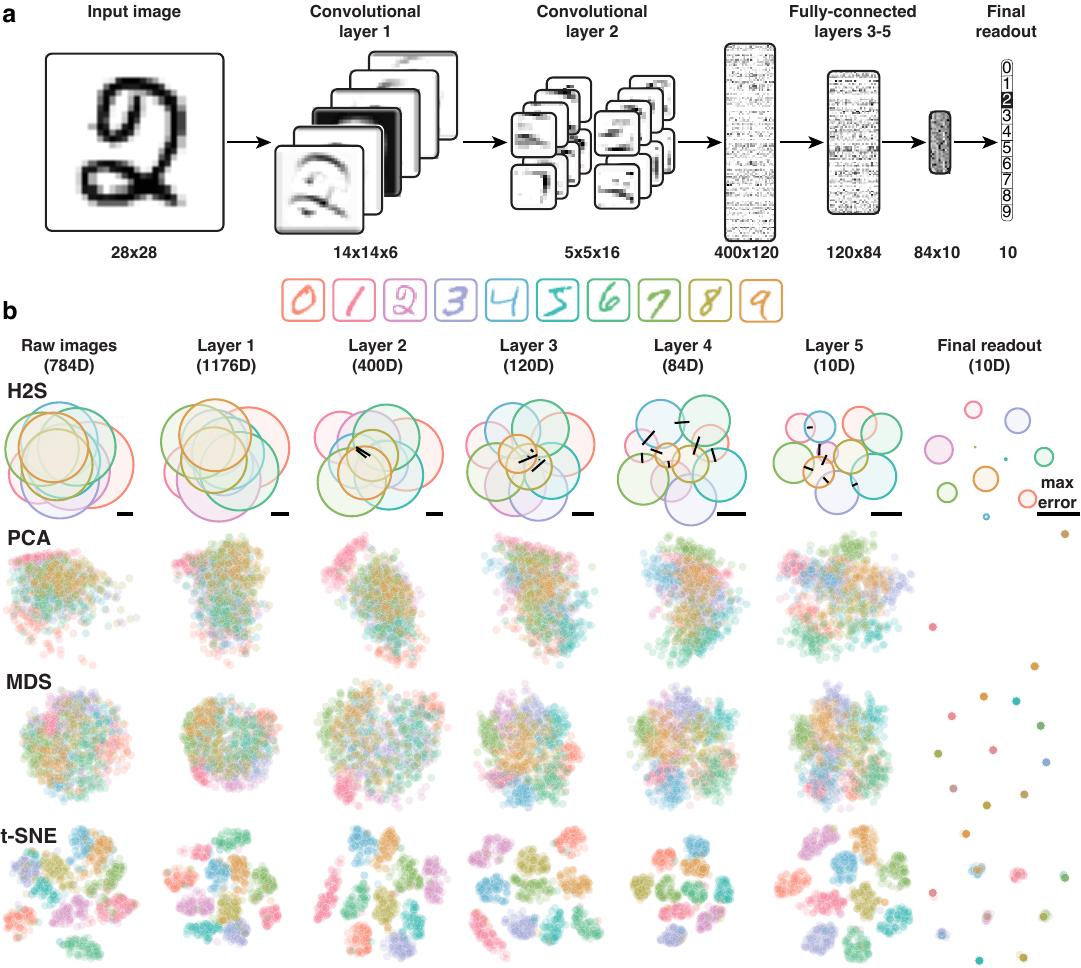}
  \vspace{4mm}
  \caption{\label{fig:lenet}\textbf{H2S best captures the gradual separation of handwritten digit representations in successive layers of a neural network.} (\textbf{a}) We trained a simple variant of the LeNet-5 neural network to distinguish handwritten digits 0 through 9 in the MNIST dataset. Activations in layers 1-4 are followed by linear rectification, and layers 1 and 2 are further followed by 2x2 max-pooling (see methods for details). Pixel intensities represent activations in each layer of the neural network to the example `2' input image shown on the left. Each layer's activation pattern shown represents a single datapoint in the visualizations below. Numbers below each layer indicate the size of the activations for convolutional layers (after max-pooling), and the size of the connection matrices for fully connected layers (their vector outputs comprise the activations visualized below). (\textbf{b}) H2S, PCA, MDS, and \textit{t}-SNE visualizations of either the raw data (leftmost column), the activations of all neurons in a given layer (middle 5 columns), and the final readout of the last layer's output (rightmost column). \revadd{H2S best reveals the progressive disentangling. PCA can be understood as choosing a viewpoint that maximizes the variance in the 2D image plane, projection onto which collapses many points on top of each other. MDS reveals some progressive clustering. \textit{t}-SNE correctly shows that the digit distributions are nonoverlapping even at the image level, but fails to reveal their progressive segregation into hyperspherical domains.} The black lines in the H2S visualizations that appear over the circles indicate overlaps in the visualization that have flipped sign: they are in fact statistically significant margins in the original high-dimensional data (and vice versa for lines in visualized margins). Each of these lines' lengths indicate the size of that particular margin/overlap error. For PCA, MDS, and \textit{t}-SNE, the same random selection of 1,000 datapoints out of the full 10,000-datapoint validation set are plotted.}
\end{figure}

In the lateral occipital complex (LOC), by contrast, bodies and faces elicit response patterns falling into nonoverlapping hyperspheres, consistent with an explicit representation of object category \cite{Grill-Spector2001}. In the fusiform face area (FFA) \cite{Kanwisher1997}, the margins between the hyperspheres further increase, with faces being particularly well isolated from the other categories. The parahippocampal place area (PPA) responds selectively to scenes, only weakly to objects, and not at all to faces \cite{Epstein1998}. H2S shows that PPA response patterns fall into only slightly overlapping hyperspheres. The scene hypersphere (grey) is largest, suggesting that distinctions among scenes are more prominently represented than distinctions among images in other categories ($P < 0.001$ for scenes versus objects, and $P < 0.01$ otherwise). Similarly, body images occupy the largest hypersphere (green) in the LOC, and face images do in the FFA, suggesting that each of these areas most distinctly represents images of a preferred category.

\subsubsection*{Deep neural network (DNN) application.} %292 words
\label{ssec:lenet}

DNNs are a leading technology in computer vision and provide the best current model of primate object recognition \cite{Yamins2014,Kriegeskorte2015}. DNNs learn a large number of parameters through extensive training and acquire internal representations that are poorly understood. We used H2S to analyze how layers in a classic DNN, LeNet-5 \cite{LeCun1998}, successively ``disentangle'' categories. The network was trained to recognize handwritten digits in the MNIST data set \cite{LeCun1998mnist} (\autoref{fig:lenet}a), and achieves high accuracy on this task (99.03\% correct on 20\% held-out data).

Since the network recognizes digits with high accuracy, all layers must contain the information of what digit is present in the image. This implies that the response pattern distributions corresponding to different digits are separable by some nonlinear classifier, and are thus minimally overlapping locally in the response space. The \textit{t}-SNE visualization emphasizes nonlinear, local neighborhood relationships and accurately reveals the separate distributions in the first and all succeeding layers (\autoref{fig:lenet}b, bottom row). All layers appear similar in the \textit{t}-SNE visualization, and it is not apparent to what extent the distributions are globally entangled in the representational space of each layer. In PCA and MDS, the distributions are highly overlapping in all but the final readout layer, suggesting similar entanglement in the representations in all layers. As a result, \textit{t}-SNE, PCA and MDS fail to reveal how the representation is progressively transformed across stages of processing. H2S clearly visualizes how the different digits' territories in the model's representational spaces become increasingly separate from each layer to the next \cite{Chung2016}.

% A single model unit can perform linear classification \cite{Rosenblatt1958,duda2001} to make decisions \cite{Johnson1980} or perform object classification \cite{hung2005fast,DiCarlo2007}. Analysis of neural network representations therefore often focuses on the linear discriminability of different classes \cite{kriegeskorte2008representational,Higgins2017}. 

\begin{figure}[bh!]
  \centering
  \includegraphics[width=180mm]{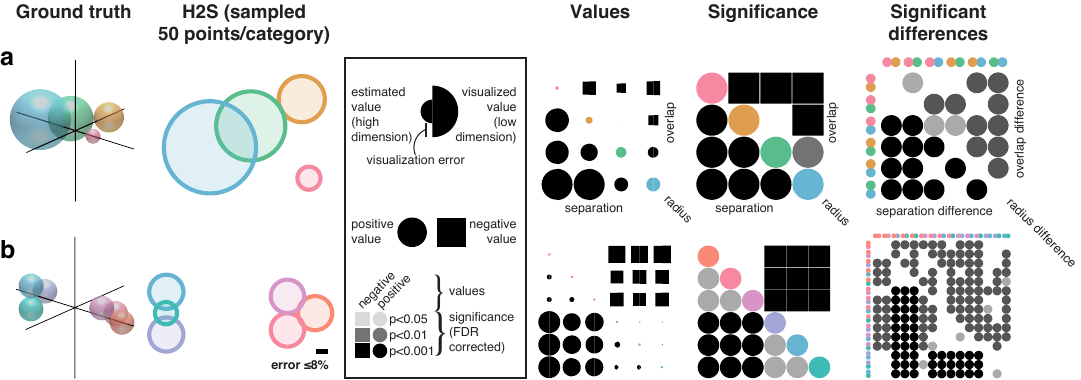}
  \caption{\label{fig:sig}\textbf{Visualizing values and statistical comparisons of sizes, separations, and overlaps.}\\
  Ground truth and H2S visualizations are shown to the left. On the right are the corresponding \textit{H2S inference diagrams}, which use elaborated forms of Hinton diagrams to visualize the summary statistics, their representation in the H2S visualization (revealing distortions due to dimensionality reduction) and inferential results. \revadd{Put another way, the summary statistic "values" visualize exactly all the quantities H2S has fit in the hypersphere fitting step, all the quantities H2S optimizes in the visualization step, and all the visualization step errors.} For both examples (\textbf{a} and \textbf{b}), three diagrams are shown, reporting the summary statistics (left), their significance (compared to 0; middle), and the significance of their pairwise differences (right). Significant separations appear in the lower triangle, overlaps/margins as circles/squares in the upper triangle, and sizes (radii) in the diagonal. To identify a specific overlap or separation, find the off-diagonal entry that is in the same row and column as the two colored circles of interest in the diagonal. Squares indicate negative values (negative overlaps, or margins), and circles indicate positive values. In the value diagram, they are split in half, with the radius of the left half indicating the estimated value in the high-dimensional space, and the right half the visualized value. The difference in the left and right half indicates the visualization error, which in these two cases, is very low (higher visualization errors can be seen in \autorefsupp{supp:lenet}). The radii of all summary statistic values are on the same scale. In the significance diagrams to the right, the darker circles indicate stronger statistical significance; the absence of a circle indicates a lack of statistical significance. (\textbf{a}) Four spheres, each a different size, all aligned to a single plane. Their summary statistics are perfectly captured by H2S, so there is no significant overall error to report (and thus the error bar is invisible). (\textbf{b}) Six equally-sized spheres, forming two groups of three equidistant spheres. In order to account for the inter-group distances, H2S introduces visualization error into some overlaps.}
\end{figure}

\subsubsection*{Nonparametric inference.} %206 words
\label{ssec:sigtest}
In addition to visualizing the size, separation, and overlap summary statistics, H2S tests these quantities and their differences for significance. The summary statistics and their differences are each evaluated using frequentist methods. They give confidence intervals and measures of significance using permutation and bootstrap methods (see \nameref{supp:methods} \nameref{supp:inference} for details). All significance tests use false discovery rate (FDR) correction to account for multiple comparisons \cite{Benjamini1995}. Furthermore, H2S computes the significance of inter-class pairwise differences between summary statistics, answering questions such as: is the distance between classes A and B significantly greater than the distance between classes B and C?

The summary statistics and inferential results are compactly visualized in \textit{H2S inference diagrams}, which use matrix-shaped, elaborated versions \revadd{of} Hinton diagrams \cite{Hinton1991} (\autoref{fig:sig}). One ($T\times T$) matrix conveys the values of the summary statistics and their representation in the H2S visualization, revealing any distortions due to dimensionality reduction. A corresponding ($T\times T$) inferential matrix conveys which summary statistics deviate significantly from 0. A third matrix ($(T^2-T)/2 \times (T^2-T)/2$) gives inferential results for pairwise comparisons among the summary statistics with the lower triangle representing the pairwise separation comparisons, the upper triangle representing the overlap comparisons, and the diagonal representing the radius comparisons (\autoref{fig:sig}). \autorefsupp{supp:lenet} represents data from \autoref{fig:lenet} in this format.

%%%%%%%%%%%%%%%%%%%%%%%%%%%%%%%%%%%%%%  DISCUSSION  %%%%%%%%%%%%%%%%%%%%%%%%%%%%%%%%%%%%%%
\section*{Discussion}
% 899 words
% The Discussion should be succinct and must not contain subheadings.

Visualization of high-dimensional data requires us to choose what information is to be conveyed. H2S is useful when we have class labels and are interested in the global relationships among the labeled distributions. The visualization conveys the relationships among a set of hyperspheres, each containing one of the labeled distributions. The enclosing hyperspheres capture the general regions occupied by the distributions, and are therefore often of interest even when the distributions are not isotropic, let alone uniform within the hyperspheres. Spheres in 3D (or circles in 2D) provide a natural visual language for conveying the hypersphere sizes, center separations, and overlaps, either exactly (for up to 4 classes in 3D or up to 3 classes in 2D) or approximately. 

An enclosing hypersphere provides a simple and intuitive characterization of the location and extent of a distribution, defined by just the center and the radius. The interpretation is particularly simple when the distribution is approximately $N$-ball uniform or Gaussian isotropic. In those cases, the sizes, separations, and overlaps of the visualization spheres directly convey the corresponding parameters of the high-dimensional $N$-ball or Gaussian distributions. However, H2S is useful to convey the relative locations and extents for any distributions. The visualization spheres, in this more general setting, must not be interpreted as conveying the shapes of the high-dimensional distributions, which may be complex and non-isotropic. Instead the spheres of the visualization convey the enclosing hyperspheres of the high-dimensional distributions, and their relationships. A margin between two visualization spheres implies that the corresponding two distributions do not overlap. Conversely, however, overlap between two visualization spheres does not imply overlap between the corresponding high-dimensional densities when these are not isotropic.

Two additional considerations make it attractive to visualize the relationships among the enclosing hyperspheres whenever the data distributions are high-dimensional. First, for many compact high-dimensional distributions, including Gaussians and hypercube uniform distributions, most points fall within a small margin of a hypersphere. An isotropic Gaussian, for example, can be characterized by its center and the radius of a hypersphere enclosing most of the distribution. The choice of the fraction of the distribution to be enclosed only slightly changes the radius in high dimensions, because the distances to center have little variance. Second, whenever the number of data points per class is smaller than the number of dimensions, it is not realistic to characterize the high-dimensional distributions in detail. This is often the case in practice. For example, the cognitive neuroscience application (\autoref{fig:trans62}) had only 12 examples per category and a 320-dimensional brain activity pattern space. In this scenario, it is impossible to estimate even an anisotropic Gaussian model of the density. An enclosing hypersphere still gives a useful sense of the relative location and extent of the distribution.

We demonstrated H2S on synthetic data (\autoref{fig:examples}), revealing its robustness to changes of data dimensionality and to imbalances in the number of data points available for different classes. PCA, MDS, and \textit{t}-SNE are all sensitive to both dimensionality and sampling imbalances because they attempt to embed the individual points, rather than to capture the relationships between the distributions (\autoref{fig:examples}). Applied to representations in the human brain and in a neural network model, H2S revealed how categories \revadd{of visual stimuli} become successively disentangled across stages of representation, while PCA, MDS, and \textit{t}-SNE did not (\autoref{fig:lenet}). %With \textit{t}-SNE, categories appear separate even in lower layers of the network, reflecting the fact that the distributions, though entangled, do not overlap in the high-dimensional space. PCA and MDS show small decreases of overlap of the distributions. However, this is difficult to interpret because the distributions of the points actually do not overlap (as \textit{t}-SNE shows) only their enclosing hyperspheres do (as H2S shows). 
Each method provides a different perspective on the data and can be essential when the statistics it reveals are of particular interest. The unique feature of H2S is that it reveals the relationships among the enclosing hyperspheres.

H2S is attractive when we are interested in the relationships among a set of distributions, of which the data points are samples. It is not appropriate when each data point has a unique meaning and we are interested in the relationships among individual data points within each class. \revadd{H2S results, like those of any other visualization method, should be interpreted in awareness of the assumptions. The user should be aware that overlapping visualization spheres do not imply overlapping distributions when those distributions are non-isotropic (see also \cite{Das2020}, for a related cautionary example in the context of connectivity inference).} By summarizing all data points in each class with a single visual primitive, H2S achieves robustness to the particular sample of points provided and to imbalances in the number of data points available for each class. 
It reveals the global relationships among distributions, rather than idiosyncrasies of a particular data set.

At a more general level, H2S introduces a novel approach to visualizing data, where each distribution is represented by a 2D or 3D solid object. The properties of the visualization objects and their arrangement convey a set of summary statistics of interest of the high-dimensional distributions. Here we used spheres as the objects, providing a particularly simple visual language. The language of spheres is naturally suited to expressing the sufficient statistics of hyperspheres. Future work could use visualization objects of a different shape class (e.g., ellipsoids) and could expand the set of summary statistics (e.g., to include covariance information). Richer summary statistics will require a more expressive visual language. H2S provides a first example of this novel approach, focusing on location and extent, two of the most basic properties of distributions.

%%%%%%%%%%%%%%%%%%%%%%%%%%%%%%%%%%%%%% BIBLIOGRAPHY %%%%%%%%%%%%%%%%%%%%%%%%%%%%%%%%%%%%%%
\bibliography{h2s-scirep} % should not exceed 40 references

%%%%%%%%%%%%%%%%%%%%%%%%%%%%%%%%%%%%%%      ETC     %%%%%%%%%%%%%%%%%%%%%%%%%%%%%%%%%%%%%%
% \section*{Acknowledgments (not compulsory)}

% Acknowledgments should be brief, and should not include thanks to anonymous referees and editors, or effusive comments. Grant or contribution numbers may be acknowledged.

\section*{Author contributions statement}

NK and ADZ conceived and designed the method. ADZ conducted the neural network simulations. AW conducted and ADZ and AW analyzed the functional MRI experiment. ADZ created software. ADZ and NK wrote the manuscript.

\section*{Additional information}

The authors declare no competing interests.

\newpage
%%%%%%%%%%%%%%%%%%%%%%%%%%%%%%%%%%%% ONLINE METHODS %%%%%%%%%%%%%%%%%%%%%%%%%%%%%%%%%%%%%
\section*{Methods}
\label{supp:methods}

\subsubsection*{Convergence of Gaussian, uniform $N$-ball, and hyperspherical distributions in high dimensions.}
\revadd{Gaussian distributions are ubiquitous. Visualizing a high dimensional isotropic Gaussian in 2D or 3D can be done by displaying a circular or spherical level-set. This fact that we are using spheres as our primary visualization element leads to a natural next question: what if we use hyperspheres, as opposed Gaussians, for example, to describe the data? By examining the distribution of the distances of the points to the center (mean) for different high-dimensional distributions, we in fact see that isotropic Gaussian, uniform $N$-ball, and hyperspherical distributions converge.}

\revadd{For a uniform $N$-ball, the distance-to-center (D2C) distribution is uniform for 1 dimension (where the ball reduces to an interval), and tends to a delta function as dimensionality increases (\autorefsupp{supp:histograms}a). For a uniform $N$-ball in a finite number of dimensions greater than 1, the D2C distribution has a single (leftward) tail and the mode is the maximum distance to center, which equals the radius of the ball. Similarly, for a Gaussian or hypercubic uniform distribution, the D2C distribution tends toward a delta function as dimensionality increases. However, for these two distributions, the D2C distribution has two tails (\autorefsupp{supp:histograms}b,c).}

\revadd{For the Gaussian and hypercubic uniform distributions, the D2C distributions have modes that grow monotonically with dimensionality. For the Gaussian, the support of the D2C distribution extends to positive infinity. The enclosing hypersphere would thus be infinite and uninterpretable. Since the Gaussian and uniform hypercubic D2C distributions both still tend toward a delta function, using the mode of the D2C distribution as the radius parameter to visualize in H2S appears to be the best choice: in finite dimensions, most D2C fall within a small margin of the mode.}

\revadd{These observations, and the fact that the shape of the D2C distribution is in fact Gaussian (\autorefsupp{supp:histograms}c), is a consequence of the Central Limit Theorem (CLT). By the CLT, the sum of squared deviations across $N$ dimensions concentrates around its mean, explaining both the Gaussian shape of the D2C distribution and its decreasing variance with dimensionality.}\\

\subsubsection*{Hypersphere parameter estimation from high-dimensional data.}
\revadd{By default, we use the 95th percentile D2C estimators (see \nameref{sssec:d2c_estimators}). Combining the observation that the distributions converge in high dimensions, with our intuition of visualizing the level set of a Gaussian, led us to explore several different estimators that each make different assumptions about these three underlying distribution types.}

\revadd{We began by estimating the parameters of enclosing $N$-balls directly. For various reasons, including accuracy, assumptions about the underlying distribution types which may or may not be possible, and computational efficiency, we explored several alternative estimators. This section describes all these different estimators, and the following section compares their performance on synthetic data.}

\begin{figure}
  \centering
  \includegraphics[width=130mm]{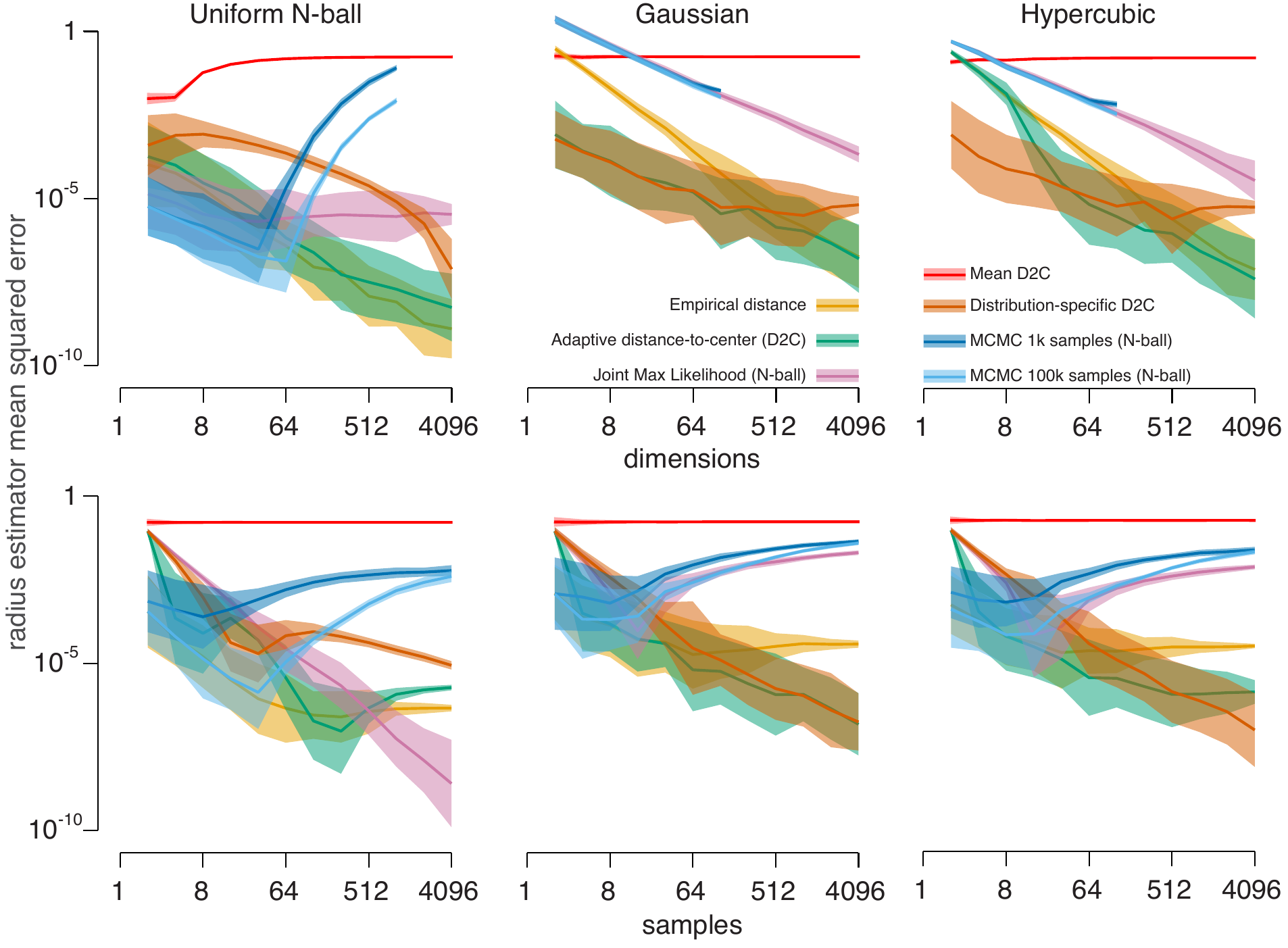}
  \caption{\textbf{H2S radius estimator performance.}\\
  Comparison of estimator accuracy (radius-normalized squared error) for radii, on uniform $N$-ball (left), Gaussian (center), and uniform hypercube (right) generating distributions. In each case in the top row, 200 points were randomly drawn, and re-drawn 100 times to generate the bootstrapped confidence intervals indicating $\pm1$ standard deviation. The bottom row shows the radius estimator error as a function of samples, for 200-dimensional data. Estimators that use a $N$-ball distributional assumption are indicated in the legend. MCMC estimators were not computed for high dimensions due to computational constraints.
  \\
  % \TODO{Could redo with more bootstraps, $\pm1$ s.e.m. More experiments: error vs added noise}
  }\label{supp:performance}
\end{figure}

Estimation of the parameters of \revadd{a }minimum enclosing ball, the center and radi\revadd{us}, is a well studied problem \cite{Kumar2003,Badoiu2003,Fischer2003,Ding2019}. Any of the estimators offered in the literature can be used in the context of H2S. Here we implemented a maximum-likelihood (ML) estimator, a Bayesian Markov-Chain-Monte-Carlo (MCMC) estimator, and several heuristic estimators that use summary statistics of the distribution of the distances from individual data points of a class to the center (estimated as the mean) and of the distribution of pairwise distances. All estimators are described below, along with an empirical comparison on simulated data. In brief, theoretical considerations and simulations showed the following: the ML estimator overfits and underestimates the radius. The Bayesian MCMC estimator is computationally intensive and not optimally stable in practice. This motivates consideration of heuristic estimators. The mean provides a good estimate of the center and a good basis for estimating the radius from the distribution of distances between the individual points and the center estimate. Simulations with data generated from Gaussian, uniform $N$-ball, and uniform hypercubic distributions suggest the use of the adaptive radius estimator $\hat{r}_{adapt}$ (\autoref{eqn:deviation}), which works robustly across distributions, dimensionalities, and numbers of samples, and is fast to compute \revadd{(}\autorefsupp{supp:performance}\revadd{)}.\\

\begin{figure}
    \centering
    \includegraphics[width=130mm]{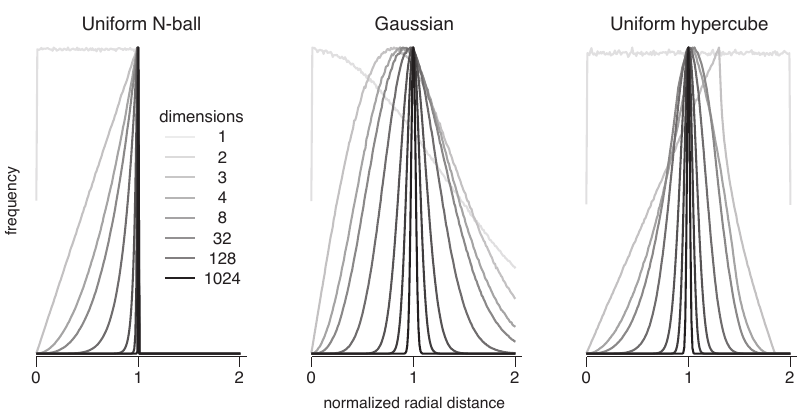} % 100,000 points
  \caption{\textbf{Different distributions tend towards the hyperspherical in high dimensions.}\\
  Histograms of 100,000 sample points' distances from the center of a uniform ball (left), Gaussian distribution (center), and uniform hypercube (right) of increasing dimensionality (increasingly dark lines). All histograms are vertically scaled so that their maximum frequencies match. The Gaussian distribution radial distances are normalized to $\gamma(N)$ (see \autoref{eqn:gamma}). The uniform hypercube radial distances are normalized to the expected mean distance from the center, given by $\int_{[-\frac{1}{2},\frac{1}{2}]^N} ||\mathbf{x}||_2 \,d\mathbf{x}$, which quickly converges to $\sqrt{N/12}$ with increasing $N$. The distribution types converge in high dimensions to a delta function at 1: the distribution for a uniform unit hypersphere.}\label{supp:histograms}
\end{figure}

\paragraph{Maximum-likelihood estimator.}
\revadd{The} first estimator is the maximum-likelihood (ML) estimate of a uniform $N$-ball distribution, which Gaussian, uniform $N-$ball, and hyperspherical distributions converge towards in high dimensions (\autorefsupp{supp:histograms}). This estimator finds the smallest (i.e. highest density) ball that contains all data points. The parameters are estimated by minimizing the maximum distance to center across all points. The minimum hypersphere's center $\hat{\mathbf{c}}_{ML}$ and radius $\hat{r}_{ML}$ are:
\begin{align}
  \hat{\mathbf{c}}_{ML} = \argmin_{\mathbf{c}} \; \max_i\: \|\mathbf{X}_i - \mathbf{c}\|_2\\
%\end{align}
%\begin{align}
  \hat{r}_{ML} = \; \max_i\: \|\mathbf{X}_i - \hat{\mathbf{c}}_{ML}\|_2
\end{align}

Finding the smallest hypersphere requires iterative optimization. The ML estimator finds the smallest possible hypersphere and is therefore certain to underestimate the actual radius of the data-generating $N$-ball distribution.\\

\paragraph{Bayesian (MCMC) estimator.}
The shortcomings of the ML estimator motivate a Bayesian approach, where we infer the full posterior over $N$-ball parameters given the data. We can use Markov Chain Monte Carlo (MCMC) sampling to estimate the joint posterior for center and radius, using infinite, flat (improper) priors for both parameters. The log likelihood is
\begin{align}
  L({\bf X} | {\hat{\mathbf{c}}},\hat{r}) = 
  \begin{cases}
      -\infty & \text{if any } \hat{\delta}_i > \hat{r},\\
      -P \log \left( \hat{r}^N V_1(N) \right) & \text{otherwise}
  \end{cases}
\end{align}

\noindent where ${\bf X}$  is the $P$-by-$N$ matrix of data points (where $P$ is the number of points), $\hat{\mathbf{c}}$ ($N$-vector) is the center estimate and $\hat{r}$ (scalar) is the radius estimate. $V_1(N)$ is the volume of a unit-radius $N$-ball.
% \begin{align}
%     V_1(N) = \frac{pi^{n/2}}{\Gamma(\frac{n}{2})+1}
% \end{align}
The distance of each point to center, $\hat{\delta}_i$, is given by:
\begin{align}
  \hat{\delta}_i = ||{\bf X}_i - \hat{\mathbf{c}}||_2
\end{align}

\begin{figure}
  \centering
  \includegraphics[width=130mm]{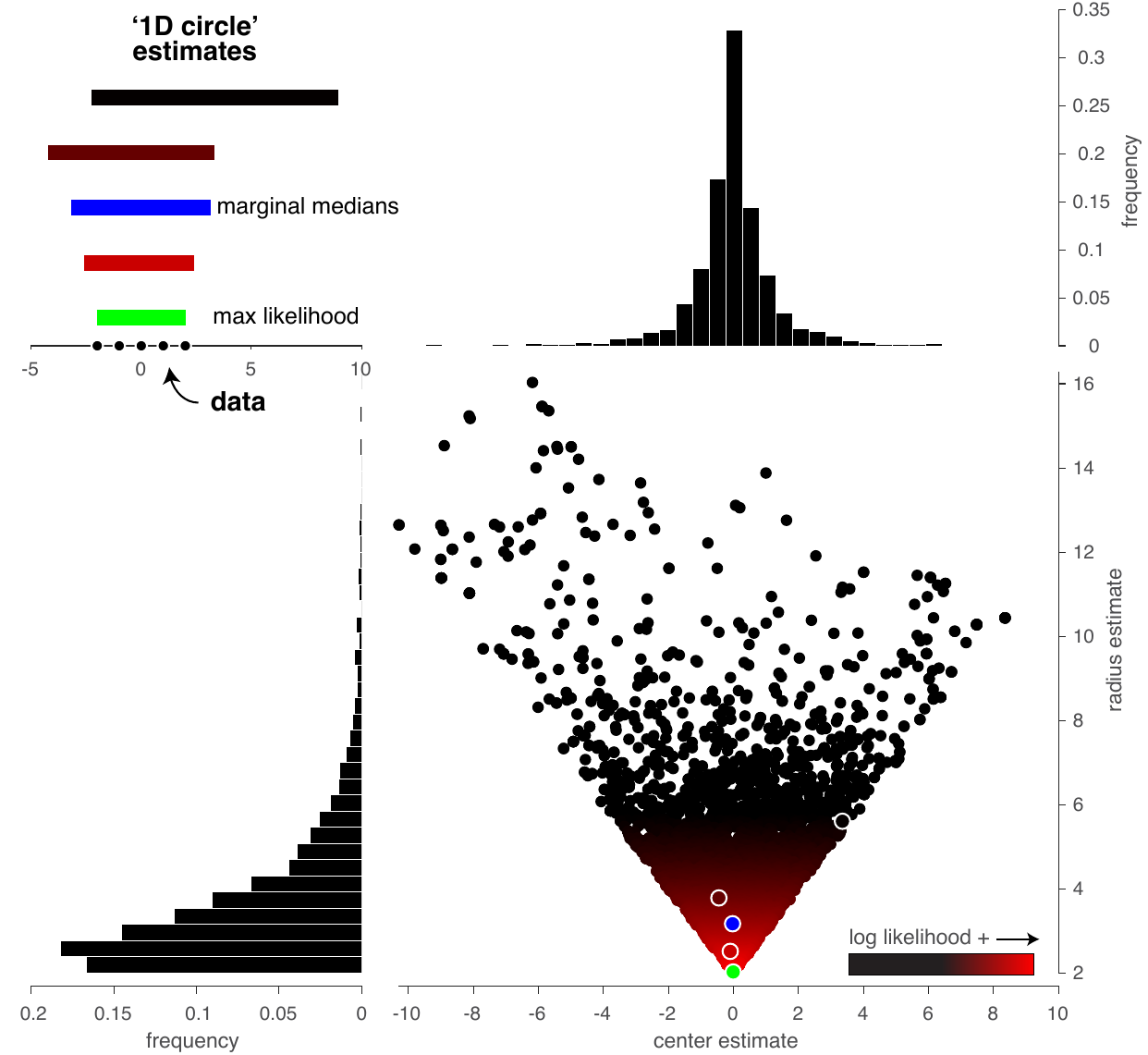}
  \caption{\textbf{H2S radius estimator performance.}\\
  A toy example: ``1D circle,'' or a closed interval on a line (top left). Black points indicate data points and line segments indicate estimated model 1D circles, colored by their likelihood. (bottom right) An example MCMC simulation with 10,000 samples (closed circles), as a function of the estimated center and radius of the interval, going from black (low log likelihood) to red (high likelihood). The open white circles indicate the sample estimates shown in top left plot. The maximum likelihood solution is shown in green. If an infinitely flat (improper) prior is assumed, the final estimate, shown in blue, is given by the median of the marginal probabilities, shown as frequency histograms.}\label{supp:mcmc}
\end{figure}

\noindent The joint posterior over center and radius is computed by MCMC sampling (\autorefsupp{supp:mcmc}). As a point estimates of the radii and inter-center distances for the H2S visualization, we use the medians of the marginal posteriors of these parameters. The marginal medians have an equal probability of being larger or smaller than the true parameters of the data-generating $N$-balls.\\

\paragraph{Distance-to-center-based estimators.}
\label{sssec:d2c_estimators}
\revadd{As we observed earlier (\autorefsupp{supp:histograms}), as dimensionality grows, the mode of the D2C distribution converges to the radius of the hypersphere that all data converge to. The mode of the D2C distribution therefore provides a good target for our radius estimator, whether the underlying distribution is a uniform ball, an isotropic Gaussian, or even a non-isotropic distribution such as a uniform hypercube. A possible alternative threshold for visualization is the 95th percentile of the D2C distribution, ensuring that 95\% of the points fall into the hypersphere. The specific percentile is somewhat arbitrary. However, this choice makes little difference for the radius estimate, because the D2C distribution has small variance for high-dimensional data.}

The \revadd{first} question then is how to estimate the mode of the D2C distribution. For the high-dimensional ($N>100$) case, the D2C distribution is very concentrated, so the mean or the median of the empirical D2C distribution could be used as an estimate of the mode, where the center is estimated as the mean of the data points. This, however, leads to negatively biased estimates of the radius parameter (\autorefsupp{supp:performance}), because using the centroid overfits the data and minimizes the sum of the squared D2C. Moreover, we would like our estimator to work even for lower-dimensional data, where the mean or the median of the empirical D2C distribution is not a good estimate of the mode of the population D2C distribution. Since the mode of the D2C distribution depends on the underlying distribution, we developed radius estimators that assume the distributions are of a particular shape. These estimators use summary statistics of the empirical D2C distribution and provide fast and accurate radius estimates (\autorefsupp{supp:performance}).

Under the uniform $N$-ball assumption, the mode of D2C distribution is the maximum distance to center (\autorefsupp{supp:histograms}). The D2C-based radius estimate for a uniform $N$-ball is
\begin{align}
  \hat{r}_{DCB1} = (1 + P^{-N})\max_{i\in \{1..P\}}\hat{\delta_i} 
\end{align}
where $\max_i \hat{\delta_i}$ is the empirical maximum distance to the mean of the points and the factor preceding it corrects the estimate upward to account for the fact that the maximum empirical distance to centroid will be smaller than the radius.
%\textbf{CUT?[ } This estimator attempts to generalize, to multiple dimensions, the solution to the German Tank problem.\textbf{ ]CUT?}

For a unit-variance Gaussian distribution, the D2C distribution is the chi distribution, and its mean $\gamma(N)$ is a function of the dimension $N$:
\begin{align}
\label{eqn:gamma}
  \gamma(N) = \sqrt{2}\frac{\Gamma((N+1)/2)}{\Gamma(N/2)}
\end{align}
where $\Gamma(z) = \int_0^\infty x^{z-1}e^{-x}\;dx$ is the Gamma function. The D2C-based radius estimator for a Gaussian scales the standard deviation of the point coordinates to this mean D2C defined by $\gamma(N)$:
\begin{align}
  \hat{r}_{DCG}  = \gamma(N) \sqrt {{\frac{1}{N(P-1)}}\sum _{i=1}^{P}\hat{\delta}_i^2}.
\end{align}
\begin{comment}
ALTERNATIVE
\begin{align}
  \hat{r}_{DCG} = \gamma(N) \cdot \sqrt {{\frac {1}{PN-N}}\sum _{i=1}^{P}\sum _{j=1}^{N}(\mathbf{X}_{ij}-\mathbf{c}_j)^{2}} = \gamma(N) \cdot \sqrt {{\frac {1}{N(P-1)}}\sum _{i=1}^{P}\hat{\delta}_i^2}
\end{align}
\end{comment}

For a uniform hypercubic distribution, the expected mean distance from the center is
\begin{align}
\int_{[-\frac{1}{2},\frac{1}{2}]^N} ||\mathbf{x}||_2 \,d\mathbf{x},
\end{align}
which quickly converges to $\sqrt{N/12}$ with increasing $N$. We could use an estimate of the width $w$ of the hypercubic distribution and estimate the radius as $w \sqrt{N/12}$.  However, we instead define the D2C-based estimator for hypercubic distributions, $\hat{r}_{DCC}$, as the median of the empirical D2C distribution, because simulations show that the median is more stable.\\
% \begin{align}
%   DCE_C = \underset{i\in \{1..P\}}{\mathrm{median}} (\hat{\delta_i}) 
% \end{align}

\paragraph{Adaptive radius estimator.}
The D2C-based estimators require the user to specify an assumed shape of the distribution. Since the purpose of dimensionality-reducing visualization is often exploratory, it is desirable to have an option that works robustly and does not require the user to assume any particular distribution. Based on empirical analyses of the D2C distributions for various multivariate distributions, we developed a fast, general-purpose radius estimator that uses the variance of the empirical D2C distribution to decide whether the distribution is more like a Gaussian (with infinite support and a wider D2C distribution), or more like a ball or hypercube (with finite support and a narrower D2C distribution).

\begin{figure}
  \centering
  \includegraphics[width=180mm]{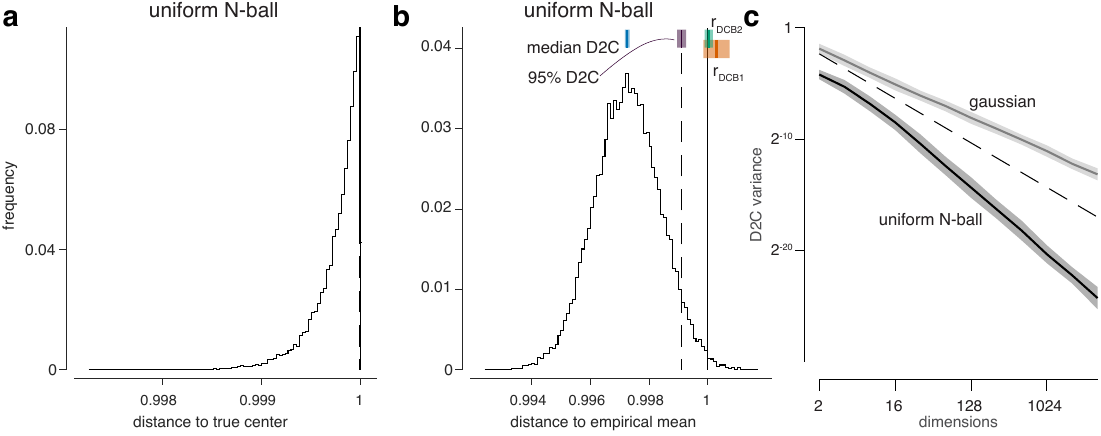}
  \caption{\textbf{Adaptive D2C estimator motivation.}\\
  (\textbf{a}) Histogram of distances-to-center (D2C) for a uniform $N$-ball ($N=4096$) relative to the ground-truth center, combined over 100 independent simulations of a data set with 200 sample points. This corresponds to the right-most condition plotted in the top-left of \autorefsupp{supp:performance}. (\textbf{b}) Histogram of the D2C for the same simulated data as in (a), but computed relative to the empirical means (centroids) in each of the 100 simulated data sets, as would happen in data analysis, where the center must be estimated. The empirical median D2C and its $\pm1$ standard error is plotted above in blue. The empirical median D2C is a negatively biased estimate of the radius. The 95th percentile estimator, enclosing 95\% of points (the dashed line), is a fast and closer approximation of the true radius. The estimators $r_{DCB1}$ and $r_{DCB2}$ aim to reduce the bias. $r_{DCB1}$ slightly overestimates the true radius, whereas $r_{DCB2}$ is more accurate. The $r_{DCB2}$ estimator corrects the negative bias by adding a small multiple $\xi(N)$ (estimated by simulation for each $N$) of the standard deviation of the empirical D2C distribution (\autoref{eqn:rdcb2}). (\textbf{c}) D2C variance decreases with increasing dimensionality. The uniform $N$-ball and Gaussian distributions are linearly separable in a log-log plot of variance of the D2C distribution versus dimensionality. The dashed line corresponds to the criterion used in $r_{adapt}$ (\autoref{eqn:deviation}) to discriminate Gaussian and uniform $N$-ball distributions.
  }\label{supp:radapt}
\end{figure}

\revadd{The variance of the D2C distribution provides a theoretically motivated criterion for this selection: by the CLT, the squared distance to center is a sum of $N$ terms whose concentration rate differs systematically between Gaussian and uniform ball distributions, making them linearly separable in log-log variance versus dimensionality (\autorefsupp{supp:radapt}).}

If the empirical D2C distribution has high variance, the radius is estimated as the $\hat{r}_{DCG}$ (aiming to estimate the mean of the D2C distribution for a Gaussian). If the empirical D2C distribution has low variance, then the median of the empirical D2C distribution is used as a basis and corrected upward by a fraction $\xi(N)$ of the standard deviation of the D2C distribution:
\begin{align}
    \label{eqn:rdcb2}
    \hat{r}_{DCB2} = \mathrm{median} (\hat{\delta_i})  + \mathrm{std} (\hat{\delta}_i)\xi(N)
\end{align}

\noindent The scaling factor $\xi(N)$ is based on simulations where data are generated from uniform $N$-ball distributions:

\begin{table}[h]
    \centering
    \begin{tabular}{|c| c c c c c c c c c c c c|}
        \hline
        $N$ & 2 & 4 & 8 & 16 & 32 & 64 & 128 & 256 & 512 & 1024 & 2048 & 4096 \\
        \hline
        $\xi(N)$ & 1.2733 & 1.0115 & 0.8796 & 0.8107 & 0.8384 & 0.8638 & 0.9579 & 1.0403 & 1.1938 & 1.4268 & 1.8384 & 2.4485 \\
        \hline
    \end{tabular}
    \caption{Simulation-derived values of $\xi(N)$. For other $N$, $\xi(N)$ is computed via linear interpolation.}
    \label{supp:table:xi}
\end{table}

The adaptive radius estimator ($\hat{r}_{adapt}$) automatically selects whether to use $\hat{r}_{DCG}$ or $\hat{r}_{DCB2}$ on the basis of the observation that the variance of D2C for both distributions follow linearly separable power laws (\autorefsupp{supp:radapt}):
% \begin{align}
% \label{eqn:deviation}
%   DEV  = 
%   \begin{cases}
%     DCE_G & \text{if } \log_2 \left(\mathrm{var} \frac{{\bf \hat{\delta}}}{\mathrm{median}({\bf \hat{\delta}})} \right) + \frac{4}{3} \log_2 \left(N \right) > -1 \;\; \text{ (Gaussian assumed)}\\
%     DCE_C + \mathrm{std} (\hat{\delta})\xi(N) & \text{otherwise}
%   \end{cases}
% \end{align}
\begin{align}
\label{eqn:deviation}
  \hat{r}_{adapt}  = 
  \begin{cases}
    \hat{r}_{DCG} & \text{if } \mathrm{var} \frac{{\bf \hat{\delta}_i}}{\mathrm{median}_i({\bf \hat{\delta}_i})}  > 2^{-1- \frac{4}{3} \log_2 N } \;\; \text{ (Gaussian assumed)}\\
    \hat{r}_{DCB2} & \text{otherwise}
  \end{cases}
\end{align}

% \underset{i\in \{1..P\}}{\mathrm{median}} (\hat{\delta_i})  + \mathrm{std} (\hat{\delta}_i)\xi(N) & \text{otherwise}

\noindent The D2C variance criterion in \autoref{eqn:deviation} distinguishes between Gaussian and $N$-ball data-generating distributions with maximum margin. This criterion also works well for hypercubic uniform distributions.\\ 
%Note that the standard deviations of the center distances decrease as a function of $\log_2 (N)$ for all distribution types, but this quantity remains relatively larger for long-tailed (Gaussian) distributions.\\

\paragraph{Pairwise-distance-based estimator.}
We can also estimate the radius of a ball distribution on the basis of an estimate of the expected distance between randomly drawn points. For a given dimensionality $N$, the expected inter-point distance $E[d_{i,j}]$ is proportional to the radius: $r=\zeta(N)\cdot E[d_{i,j}]$, where $\zeta(N)$ converges to $1/\sqrt{2}$ for large $N$. The expected-distance estimator $\hat{r}_{dist}$ uses the mean empirical inter-point distance as a basis for estimating the expected inter-point distance of the data-generating uniform $N$-ball distribution. It uses the proportionality constant $\zeta(N)$ to adjust the empirical mean inter-point distance, so as to estimate the radius:
\begin{align}
    \label{eqn:rdist}
    \hat{r}_{dist} = \zeta(N)\frac{2}{P(P-1)} \sum_{i=1}^{P-1} \sum_{j=i+1}^P ||{\bf X}_i - {\bf X}_j||_2
\end{align}

This estimator performs well and is quite robust to changes of the shape of the distribution (\autorefsupp{supp:performance}). It is useful in particular when we are only given the pairwise distances among samples, and not the point locations.

\begin{table}[h]
    \centering
    \begin{tabular}{|*p{30pt}|^p{20pt}^p{20pt}^p{20pt}^p{20pt}^p{20pt}^p{20pt}^p{20pt}^p{20pt}^p{20pt}^p{20pt}^p{20pt}^p{20pt}^p{20pt}^p{20pt}|}
        \hline
        $N$ & 1 & 2 & 4 & 8 & 16 & 32 & 64 & 128 & 256 & 512 & 1024 & 2048 & 4096 & 4097+ \\
        \hline
        $1/\zeta(N)$ & 0.6673 & 0.9039 & 1.1043 & 1.2407 & 1.3230 & 1.3657 & 1.3898 & 1.4020 & 1.4081 & 1.4111 & 1.4127 & 1.4134 & 1.4138 & $\sqrt{2}$ \\
        \hline
    \end{tabular}
    \caption{Experimentally-derived values of 1/$\zeta(N)$. For other $N$, $\zeta(N)$ is computed via linear interpolation.}
    \label{supp:table:zeta}
\end{table}

\subsubsection*{Comparison of the estimators on simulated data.}
We compared the performance and speed of the different estimators on synthetic data---including uniform $N$-balls and hypercubes, and Gaussians---as a function of dimensions and number of samples (\autorefsupp{supp:performance}). In practice, MCMC performs well for uniform $N$-ball distributions, and increasing the number of MCMC samples only significantly improves performance when it is significantly higher than the data dimensionality. For both the uniform and Gaussian distributions, however, the respective D2C-based estimators ($\hat{r}_{DCB}$, $\hat{r}_{DCG}$, or $\hat{r}_{DCC}$, depending on the data-generating distribution; orange in \autorefsupp{supp:performance}) perform quite well, and are faster to compute. The D2C-based estimators are very accurate and fast to compute when they match the underlying distributions, but fail badly when there is a mismatch (not shown). The adaptive D2C-based estimator $\hat{r}_{adapt}$ (\autoref{eqn:deviation}) is fast to compute, does not require user input (of distributional assumptions), and performs robustly across the three data-generating distributions. It is therefore the default estimator in our current H2S implementation.
% \TODO{show a plot of this and add to table? state that automatic DCE is the default at the top of this section? may need to explain further.}

\subsubsection*{Low-dimensional sphere embedding.}
\label{supp:embed}
For the second step in the H2S algorithm, the radii of the visualization spheres are always initialized to the estimated high-dimensional radii. We initialize the centers using MDS with metric stress \cite{Young1938,Torgerson1952} as the optimization criterion. For up to 4 categories visualized as spheres in 3D (or up to 3 visualized as circles in 2D), the summary statistics will be perfectly expressed at this point. Otherwise, the low-dimensional center positions and the radii are further adjusted to minimize the sum of squared errors of the visualized distances $\tilde{d}_{ij}$, margins $\tilde{m}_{ij}$, and radii $\tilde{r}_i$ (i.e. their deviations from their high-dimensional target estimates $\hat{d}_{ij}$, $\hat{m}_{ij}$, and $\hat{r}_i$):
\begin{align}
\label{eqn:generalizedobj}
  E_{\alpha,\beta} &= \sum_{i=1}^{T-1} \sum_{j=i+1}^T      ( \tilde{d}_{ij} - \hat{d}_{ij} )^2 + 
              \alpha  \sum_{i=1}^{T-1} \sum_{j=i+1}^T \left( \tilde{m}_{ij} - \hat{m}_{ij} \right)^2 + 
                \beta \sum_{i=1}^T                    \left( \tilde{r}_i    - \hat{r}_i    \right)^2
\end{align}
where the distances are a function of the centers, and the margins are a function of the distances and radii, as previously defined in \autoref{eqn:dists} and \autoref{eqn:margins}.

\autoref{eqn:generalizedobj} represents a generalized version of \autoref{eqn:simpleobj} in which the user can choose hyperparameters $\alpha$ and $\beta$ to place more weight on representing the margins and radii accurately, relative to the distances between centers. By default, each of these hyperparameters is set to 1. One could for example, set $\beta = 10$, increasing the relative accuracy of the visualized radii, at the expense of the distances and margins. It is important to note that in the limit where $\beta\to\infty$, $\tilde{\bf r}\to\hat{\bf r}$, which makes $(\tilde{\bf m}-\hat{\bf m})\to(\tilde{\bf d}-\hat{\bf d})$, so the problem exactly reduces to finding the center parameters which minimize the squared error of the distance. This is equivalent to embedding the centers using MDS with metric stress. Since the MDS computation is simple and fast, it serves as the initialization for optimizing \autoref{eqn:generalizedobj}. For this same reason, this MDS-computed initialization is presented to the user as an option to be used as the final embedding configuration, with the radius values directly copied from the high-dimensional space. This option offers perfect radius accuracy, the best possible distance accuracy, and much faster computation, at the expense of lower margin accuracy.

When optimizing the full objective, we use MATLAB's \textit{fminunc} function for unconstrained optimization using BFGS quasi-Newton gradient descent with the following analytic gradients, where $\tilde{\bf C} = \left[ \tilde{\bf c}_1 \; \ldots \; \tilde{\bf c}_m \right]$ is the matrix concatenation of the low-dimensional center coordinate estimates and $\mathcal{T} \setminus i = \{ \{1,\ldots,T\} \setminus i \}$ is the set of integers from 1 to $T$, excluding $i$:
\begin{align}
  \frac{\partial E}{\partial \tilde{r}_i}
     &= 2\beta\left( \tilde{r}_i - \hat{r}_i \right) - 2\alpha \sum_{j \in \mathcal{T} \setminus i} \left( \tilde{m}_{ij} - \hat{m}_{ij} \right)\\
  \frac{\partial E}{\partial \tilde{\bf C}}
     &= \frac{\partial E}{\partial \tilde{\bf d}} \cdot \frac{\partial \tilde{\bf d}}{\partial \tilde{\bf C}}\\
  \frac{\partial E}{\partial \tilde{\bf d}}
     &= 2 \left( \tilde{\bf d} - \hat{\bf d} \right) + 2\alpha \left( \tilde{\bf m} - \hat{\bf m} \right) \\
  \frac{\partial \tilde{d}_{ij}}{\partial \tilde{\bf c}_i}
     &= \frac{\tilde{\bf c}_i - \tilde{\bf c}_j}{\tilde{d}_{ij}}
\end{align}
Fully expanded, the gradient of the generalized objective (\autoref{eqn:generalizedobj}) with respect to the centers is:
\begin{align}
  \frac{\partial E}{\partial \tilde{\bf c}_i}
    &= 2\sum_{j \in \mathcal{T} \setminus i} \left[ \left( \tilde{d}_{ij} - \hat{d}_{ij} \right) + \alpha \left( \tilde{m}_{ij} - \hat{m}_{ij} \right) \right] 
    \frac{\tilde{\bf c}_i - \tilde{\bf c}_j}{\tilde{d}_{ij}}
\end{align}

\subsubsection*{Human imaging experiment.}
\begin{figure}
  \centering
  \includegraphics[width=180mm]{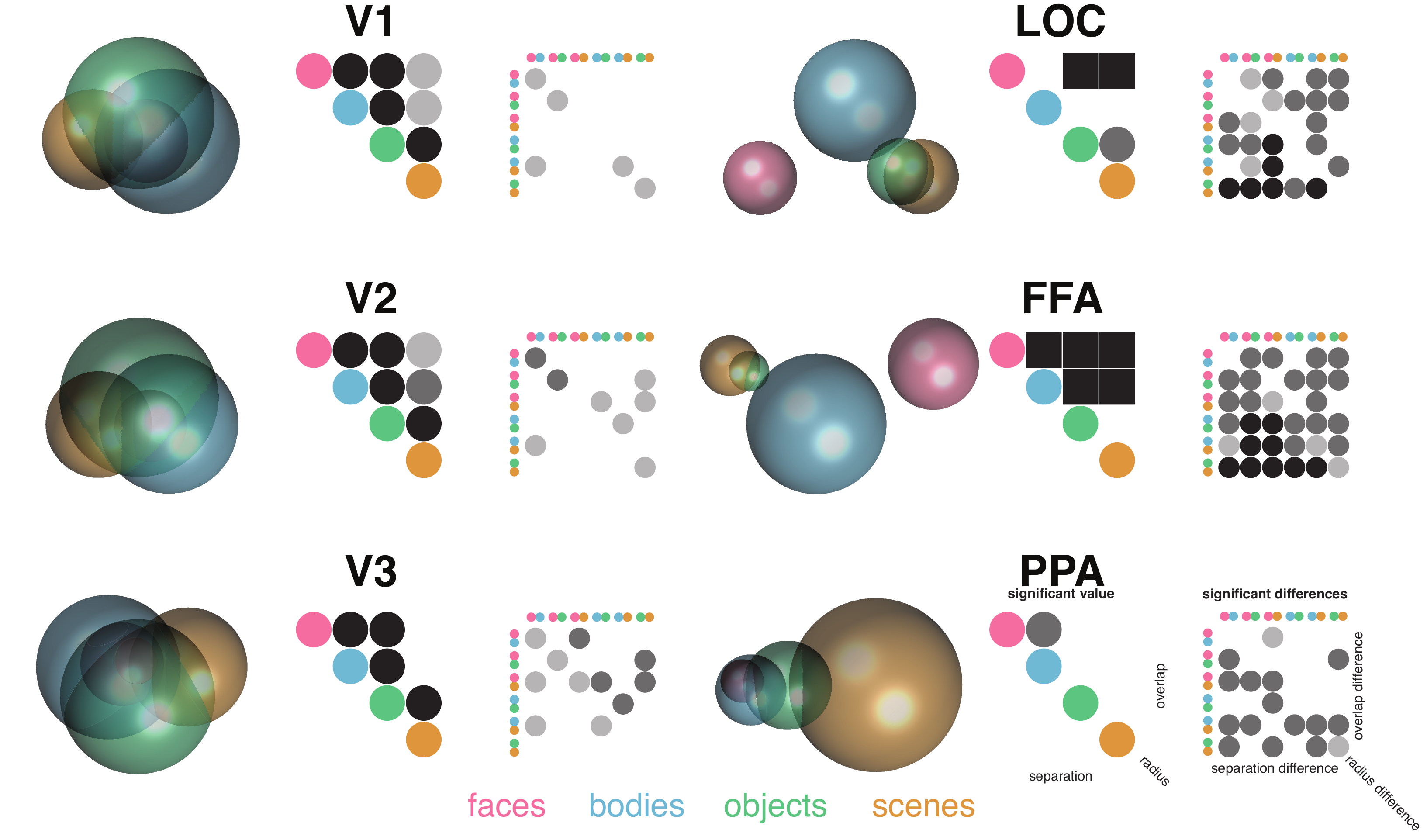}
  \caption{\textbf{H2S visualization of brain activity pattern distributions associated with visual object categories.}\\
  The top row shows the same H2S visualizations for brain regions of interest as in \autoref{fig:trans62}c.
  Modified Hinton-diagrams of the summary statistic values (second columns) and significance (third columns) show how categories are not distinguishable (in terms of overlap and separation) in early visual areas (V1-V3) but are in later areas (LOC, FFA, PPA).}\label{supp:trans62}
\end{figure}

In an fMRI experiment, 24 human subjects were presented with 48 images from visual categories such as animate/inanimate, face/body, and animal/human face (\autoref{fig:trans62}a; see \cite{Walther2015} for experimental details).
% * <adz213@nyu.edu> 2018-05-25T02:44:42.642Z:
% 
% I know there are 62 in the dataset, but I think the image-processed human faces were left out of the H2S visualization.
% 
% ^.
Regions of interest comprising 320 voxels were then defined for six visual areas (\autoref{fig:trans62}b) using separate localizer experiments and retinotopic mapping stimuli. For each pair of stimuli, the crossnobis distance estimator \cite{Kriegeskorte2007,Nili2014,Walther2016} provided an unbiased estimate of the representational distance. The crossnobis distance matrix can then be visualized using H2S or classical MDS (\autoref{fig:trans62}c, left and right respectively, in each column). \autorefsupp{supp:trans62} shows these same H2S visualizations with their accompanying inference results visualized in H2S inference diagrams.

\subsubsection*{Deep neural network analysis.}
\begin{figure}
  \centering
  \includegraphics[width=180mm]{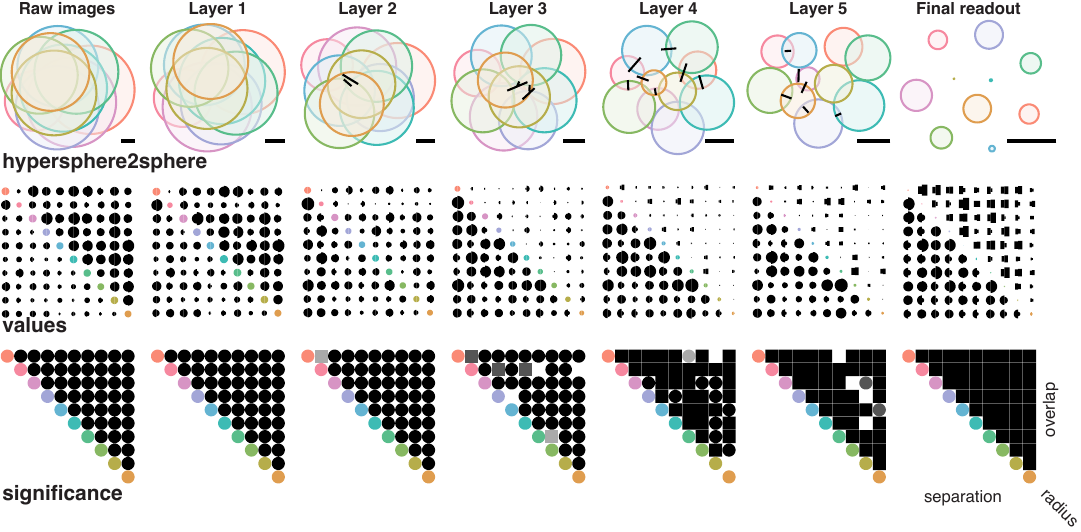}
  \caption{\textbf{H2S visualization of LeNet-5 neural network layer-wise representations of MNIST digits.}\\
  The top row shows the same H2S layer-wise visualizations as in \autoref{fig:lenet}b. Significance values computed with 10,000 bootstraps (and 10,000 jackknives for BCa-based tests). Modified Hinton-diagrams of the summary statistic values (middle row) and significance (bottom row) further make clear the gradual shift from positive overlaps to negative ones (positive margins) for representations in increasingly high layers in the network.}\label{supp:lenet}
\end{figure}

We trained a simplified variant of LeNet-5\cite{LeCun1998} on the MNIST handwritten digit dataset\cite{LeCun1998mnist}. The 28x28 MNIST images were padded to be 32x32, and served as input to the network. The first two layers are convolutional with a stride of 1, with learned 5x5 filter kernels. The first layer has 6 output channels, the second layer takes a 16x16x6 input and outputs an 8x8x16 response. Both convolutional layers are followed by 2x2 max-pooling. The following three layers are fully connected: 400x120, 120x84, and 84x10. The final predicted class label is the argmax of the 10D feature vector. The network was trained with backpropagation using stochastic gradient descent (with momentum = 0.9 and learning rate = 0.001) using cross-entropy loss. Training ran for 20 epochs of the full training set (80\% of the entire dataset) using minibatches of size 30. Classifier accuracy on the validation set (20\% of data) was 99.03\%. \autoref{fig:lenet} shows H2S visualizations of the layer-by-layer activations of this network, alongside PCA, MDS, and \textit{t}-SNE. \autorefsupp{supp:lenet} shows these same H2S visualizations with their accompanying inference results visualized in H2S inference diagrams.

\subsubsection*{Non-parametric inference.}
\label{supp:inference}
Statistics of interest optimized in the H2S objective can be optionally assessed for statistical significance via resampling methods. Significance can be first-order (is overlap A significant?) and second-order (is overlap A significantly greater than overlap B?). Nonzero radii are, by definition, always significant on a first-order basis.

We strove to use permutation tests where possible. However, some tests required traditional or accelerated bootstrap confidence intervals (BCa) \cite{Efron1987}. All tests, by default, use 5,000 resamplings (except for the BCa test which requires jackknife resampling, adding an additional $P$ resamplings).\\

\begin{figure}
  \centering
  \includegraphics[width=180mm]{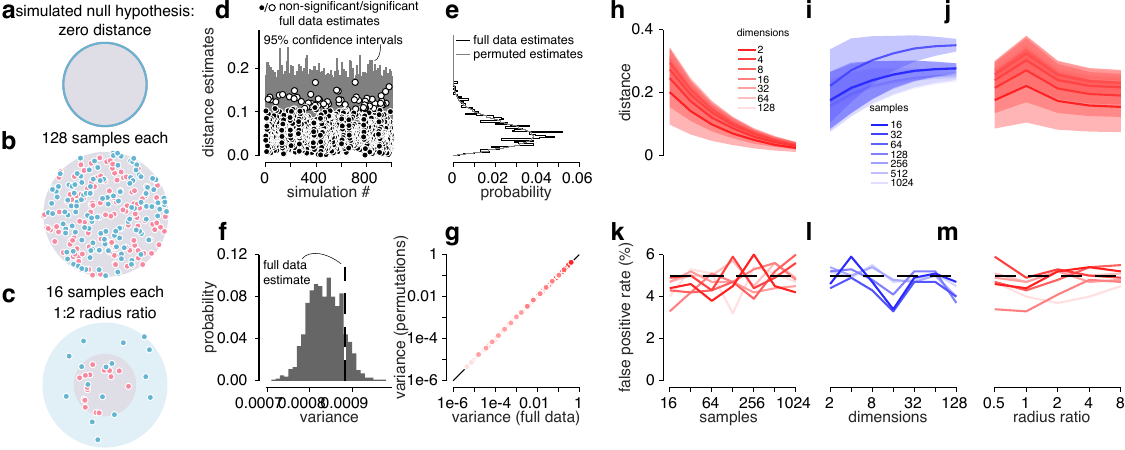}
  \caption{\textbf{Validating the one-sided permutation test for separation significance.}\\
  (a) The null hypothesis is that two hyperspheres have the same center. (b) and (c) show example simulations of the null hypothesis for different parameter choices. (d) 1,000 simulations of this null hypothesis for equal radii, 256 samples per class in 2D. The separation estimates from the full (non-permuted) data and the 0-95\% range of the permutation distribution (computed from label permutation) are shown as points and lines, respectively. White points indicate significant values.
  (e) Histograms from the 1,000 full (black) and all 5,000,000 permuted (gray) data point estimates. (f) The histogram is of each simulation's permutation distribution variance and the dashed line shows the variance of the full data estimates across all simulations. (g) The variance for permuted versus full data estimate distributions across all simulations for all parameter choices tested (lighter shades of red indicate increasing dimensionality --- see legend in (h)).
  Estimator bias is captured by the mean separation estimate across all simulations as a function of number of samples per class (h), dimensionality (i), and radius ratio (j). False positive rate for the permutation tests as a function of number of samples per class (k), dimensionality (l), and radius ratio (m).}\label{supp:statDistance}
\end{figure}

\paragraph{Separation significance.}
For each pair of classes, this one-sided test asks whether the separation between the centers of the enclosing hyperspheres is significantly larger than zero. When a distance (which is by definition nonnegative) is used to estimate the true distance from noise data, the estimate is positively biased. We therefore use an unbiased cross-validated distance estimator: the square root of the inner product of the center-difference vectors in a random split of the data \cite{Walther2016}. For a given pair of classes, the null hypothesis is that the two hyperspheres have the same center (\autorefsupp{supp:statDistance}a).Two example samplings are shown in (\autorefsupp{supp:statDistance}b-c). The cross-validated separation estimator is symmetrically distributed about 0 under the null hypothesis. We randomly permute the labels of the points between the two classes 5,000 times, re-estimate the hyperspheres, and recompute their separation. Using these permuted separations, the separation is deemed significant if it falls in the top 5\% of the permutation distribution.

We empirically validated this test for different radii and sample sizes by means of data simulated under the null hypothesis. For a given dimensionality, number of samples per class, and ratio of the radii of the two hyperspheres, we simulated data under the null hypothesis 1,000 times. The separation estimates from the full (non-permuted) data and their confidence intervals are shown as points and lines in \autorefsupp{supp:statDistance}d, respectively. Gray and black points and lines indicate non-significant and significant values --- a value computed on the full (non-permuted) data is significant when it exceeds 95\% of the permuted separations.

To verify that the permutation test distributions accurately capture the separation distribution, we compute histograms from the 1,000 full and all 5,000,000 permuted data point estimates in \autorefsupp{supp:statDistance}e (gray and black, respectively) and see that they tightly overlap. While summarizing all permutation distributions in this way is diagnostic, a further test would be to compare each permutation distribution's variance to the variance of the full data estimate distribution. \autorefsupp{supp:statDistance}f shows that the full data estimate variance is well within the orderly distribution of permutation distribution variances for each simulation. \autorefsupp{supp:statDistance}g summarizes the variance for permuted and full data estimate distributions across all simulations for parameter choices (lighter shades of red indicate increasing dimensionality --- see legend in \autorefsupp{supp:statDistance}h). Variance is very well captured, regardless of parameter choice and its resultant scale.

To examine the effect of radius ratio, dimensionality, and number of samples per class, mean estimates across all simulations are shown in \autorefsupp{supp:statDistance}h, i, and j, respectively. Unsurprisingly, estimator bias decreases for higher numbers of samples and lower numbers of dimensions. \autorefsupp{supp:statDistance}k, l, and m show the false positive rate for the permutation tests for all simulation parameter settings. Regardless of parameter choice, the false positive rate is approximately 5\%, indicating that the test well captures the statistics of the data.\\

\begin{figure}
  \centering
  \includegraphics[width=180mm]{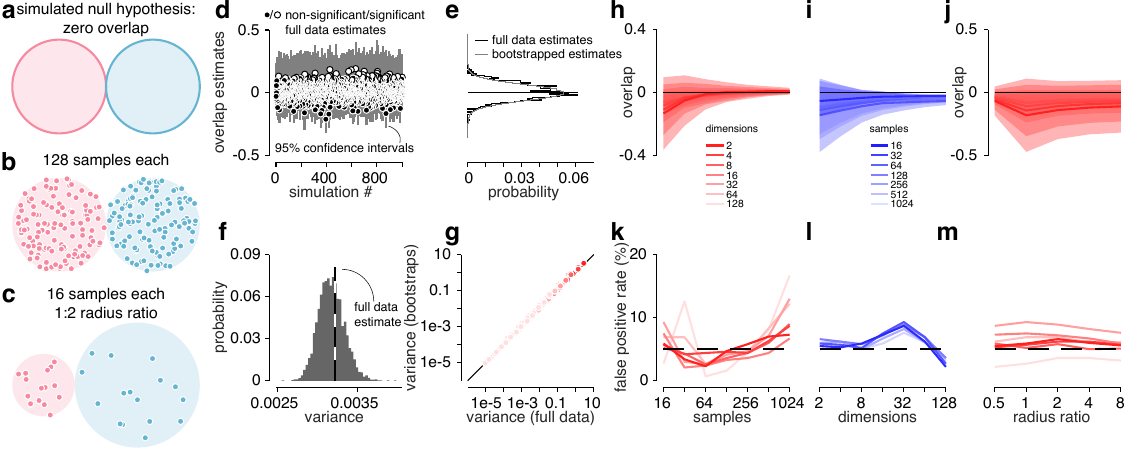}
  \caption{\textbf{Validating the two-sided accelerated bootstrap confidence interval test for overlap significance.}\\
  (a) The null hypothesis is that two hyperspheres have zero overlap. The panels in this figure follow \autorefsupp{supp:statDistance} and use the accelerated bootstrap confidence interval instead of permutation testing.}\label{supp:statOverlap}
\end{figure}

\paragraph{Overlap significance.}
For each class, this two-sided test asks whether an overlap is significantly positive or negative (negative overlaps are the same as positive margins). For a given pair of classes, the null hypothesis is that two hyperspheres have zero overlap (\autorefsupp{supp:statOverlap}a). We found while developing this test, using a permutation-based test yielded near-zero false positive rates and a classic bootstrap-based test yielded very high false positive rates, making the test overly conservative or permissive, respectively. This appeared to be due to skewness in the distribution of resampled overlaps. For a small but significant cost in computation, the accelerated bootstrap confidence interval (BCa) yields a more accurate confidence interval, as it is designed to be more robust to skewed distributions and outliers \cite{Efron1987}. For each of these 1,000 simulations, we compute bootstrap and jackknife samples of the points within each class 5,000 and $P$ times, respectively, re-estimate the hyperspheres, and recompute their overlap. Using these bootstrapped and jackknifed overlaps, we compute 95\% BCa's. The overlap estimates from the full (non-resampled) data and their confidence intervals are shown as points and lines in \autorefsupp{supp:statOverlap}d, respectively. Gray and black points and lines indicate non-significant and significant values --- a value computed on the full (non-permuted) data is significant when zero falls outside the 95\% BCa.

To verify that the BCa test distributions accurately capture the separation distribution, we compute histograms from the 1,000 full and all 5,000,000 bootstrapped data point estimates in \autorefsupp{supp:statOverlap}e (gray and black, respectively) and see that they tightly overlap. Bootstrap distributions from each simulation are centered before being combined into the single distribution shown. \autorefsupp{supp:statOverlap}f shows that the full data estimate variance is well within the orderly distribution of bootstrap distribution variances for each simulation. \autorefsupp{supp:statOverlap}g shows that variance is again well captured, regardless of parameter choice and scale.

Mean estimates show some bias \autorefsupp{supp:statOverlap}h-j, due to a tendency for the radius estimator to slightly underestimate the true radius value. Estimator bias decreases for higher dimensionality and samples in the 64-256 range. False positive rates are generally reasonably close to 5\% (\autorefsupp{supp:statOverlap}k-m), with the exception at large numbers of samples (\autorefsupp{supp:statOverlap}m). This test should only be used with caution for sample rates above 1,000 points per class.\\

\begin{figure}
  \centering
  \includegraphics[width=180mm]{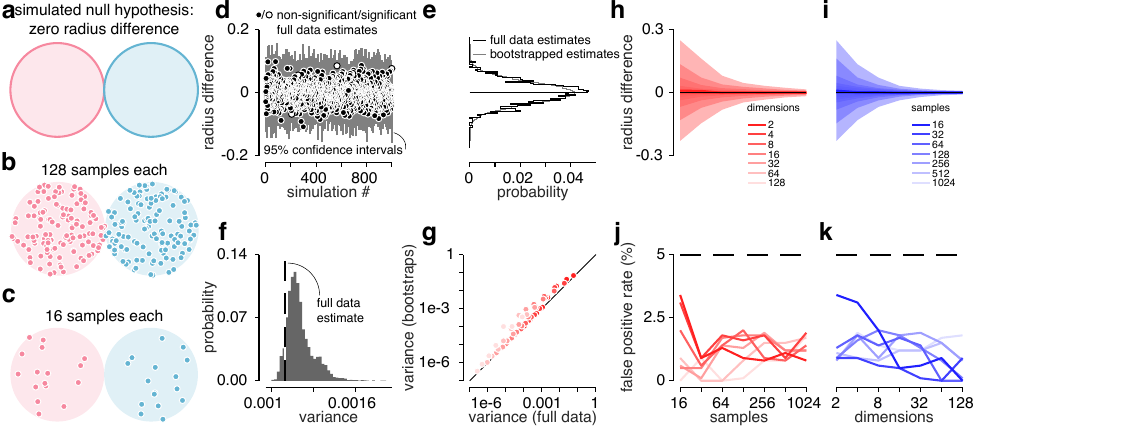}
  \caption{\textbf{Validating the two-sided bootstrap confidence interval test for radius difference significance.}\\
  (a) The null hypothesis is that two hyperspheres have the same radius (and therefore their difference is zero). The panels in this figure follow \autorefsupp{supp:statDistance} and use the accelerated bootstrap confidence interval instead of permutation testing. Radius ratio is by definition 1, so panels related to radius ratio are omitted.}\label{supp:statDiffRadius}
\end{figure}

\paragraph{Radius difference significance.}
Second-order comparisons are all concerned with whether a difference in two statistics of interest is significantly nonzero. For each class, this two-sided test asks whether one radius is significantly larger or smaller than the other (or equivalently, if the difference between two radii is significantly positive or negative). For a given pair of classes, the null hypothesis is that two hyperspheres have exactly the same radius (\autorefsupp{supp:statDiffRadius}a). For the same reasons as the overlap test, radius differences use BCa. Using 5,000 bootstrapped and $P$ jackknifed radius estimates, we compute the 95\% BCa's of their differences. A radius difference is significant when zero falls outside the 95\% BCa.

Again, the 1,000 full and all 5,000,000 bootstrapped data point estimates tightly correspond (\autorefsupp{supp:statDiffRadius}e). Bootstrap distributions from each simulation are centered before being combined into the single distribution shown. The distribution of bootstrap distribution variances is clearly non-normal (\autorefsupp{supp:statDiffRadius}f), but still the full data estimate variance is less well-centered but still well within the distribution. Bootstrap variance is slightly overestimated (\autorefsupp{supp:statDiffRadius}g), as was seen in \autorefsupp{supp:statDiffRadius}f. The effect is consistent across parameter choices, indicating a consistent behavior in the radius estimator.

Mean radius difference estimates show very little bias regardless of parameter choice (\autorefsupp{supp:statDiffRadius}h-i), indicating that the bias in the radius estimator is symmetric, and thus cancels out for the purpose of this test. Indeed, false positive rates are generally reasonably close to 5\% (\autorefsupp{supp:statDiffRadius}j-k) or are slightly higher.\\

\begin{figure}
  \centering
  \includegraphics[width=180mm]{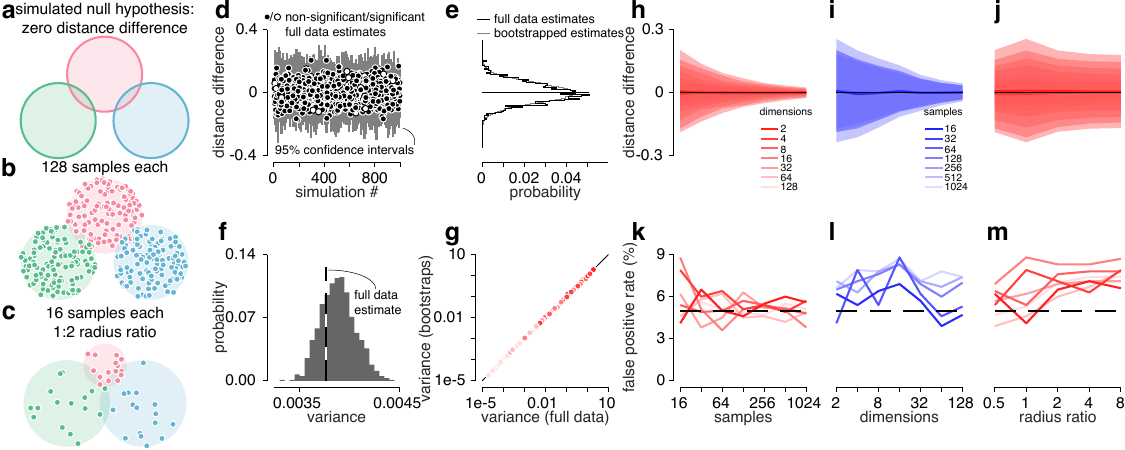}
  \caption{\textbf{Validating the two-sided bootstrap test for separation difference significance.}\\
  (a) The null hypothesis is that of three hyperspheres, two pairs have the same separation (and therefore their difference is zero). The panels in this figure follow \autorefsupp{supp:statDistance} and use bootstrap instead of permutation testing.}\label{supp:statDiffDistance}
\end{figure}

\begin{figure}
  \centering
  \includegraphics[width=180mm]{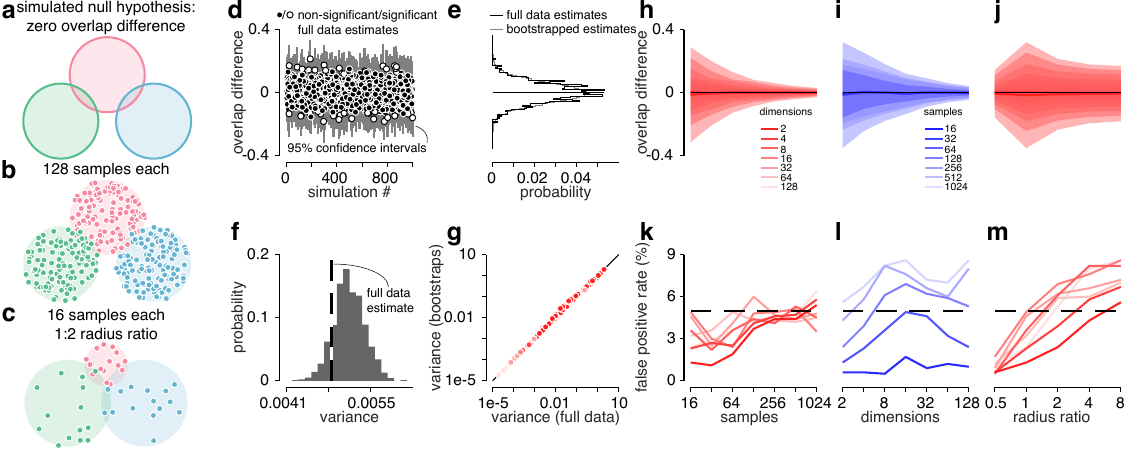}
  \caption{\textbf{Validating the two-sided bootstrap test for overlap difference significance.}\\
  (a) The null hypothesis is that of three hyperspheres, two pairs have the same overlap (and therefore their difference is zero). The panels in this figure follow \autorefsupp{supp:statDistance} and use bootstrap instead of permutation testing.}\label{supp:statDiffOverlap}
\end{figure}

\paragraph{Separation- and overlap-difference significance.} 
Separation-difference and overlap-difference tests operate in the same manner; they are two-sided bootstrap tests. These tests assess whether two overlaps (or two separations) are significantly different. For a given triplet of classes, the null hypothesis is that two hyperspheres have exactly the same separation (\autorefsupp{supp:statDiffDistance}a) or overlap (\autorefsupp{supp:statDiffOverlap}a). In these scenarios permutation testing is not applicable, so these tests use bootstrap confidence intervals. Using 5,000 bootstrapped separation/overlap estimates, we compute the 95\% bootstrap confidence intervals of their differences. A separation/overlap difference is significant when zero falls outside the 95\% confidence interval.

Again, the 1,000 full and all 5,000,000 bootstrapped data point estimates coincide well (\autorefsupp{supp:statDiffDistance}e and \autorefsupp{supp:statDiffOverlap}e). Bootstrap distributions from each simulation are centered before being combined into the single distribution shown. \autorefsupp{supp:statDiffDistance}f and \autorefsupp{supp:statDiffOverlap}f show that the full data estimate variance is well within the orderly distribution of bootstrap distribution variances for each simulation. Bootstrap variance is again slightly overestimated (\autorefsupp{supp:statDiffDistance}g and \autorefsupp{supp:statDiffOverlap}g), as was seen in \autorefsupp{supp:statDiffRadius}g. The effect is consistent across parameter choices, indicating a consistent behavior.

Mean separation and overlap difference estimates show very little bias regardless of parameter choice (\autorefsupp{supp:statDiffDistance}h-j and \autorefsupp{supp:statDiffOverlap}(h-j)). Separation difference false positive rates are generally reasonably close to 5\% (\autorefsupp{supp:statDiffDistance}k-m), but increase with increasing radius ratio. Overlap difference false positive rates are generally reasonably close to 5\% (\autorefsupp{supp:statDiffOverlap}k-m), or are slightly higher for low sample numbers.

\newpage

%%%%%%%%%%%%%%%%%%%%%%%%%%%%%%%%% SUPPLEMENTARY FIGURES %%%%%%%%%%%%%%%%%%%%%%%%%%%%%%%%%
\beginsupplement
\section*{Supplementary material}

\begin{table}[b]
  \centering
  \begin{tabular}{|*p{115pt}|^p{103pt}^p{35pt}^p{68pt}^p{43pt}^p{43pt}^p{26pt}|}
    \hline
    \rowstyle{\bfseries}
    Radius estimator & Meaning of radius & Input format & Distribution support & Low-N Accuracy & High-N Accuracy & Speed \\ \hline
    Maximum Likelihood & Expected Ball radius & Points & Ball only & High & Very high & Very slow \\ \hline
    Markov Chain Monte Carlo (MCMC) & Expected Ball radius & Points & Ball only & Very high & Very high & Slow \\ \hline
    Mean distance-to-center (D2C) & Expected radius & Points & Any & Very low & Very low & Very fast \\ \hline
    D2C-based estimators & Expected radius of \mbox{Ball}/ \mbox{Gaussian}/\mbox{Hypercube} & Points & \mbox{Ball}/\mbox{Gaussian}/ \mbox{Hypercube} only & High & High & Fast \\ \hline
    Adaptive D2C & Mean of measured \mbox{center} distances & Points & Any & High & High & Fast \\ \hline
    Pairwise distance-based empirical & Mean of measured \mbox{pairwise} point distances & Distances & Any & Medium & High & Fast \\ \hline
  \end{tabular}
  \caption{\textbf{Radius estimators compared.} The MCMC and maximum likelihood estimators jointly estimate the center and radius, while the other estimators estimate the radius directly, conditional on the maximum likelihood estimate of the center.}\label{tab:estimators}
\end{table}

\end{document}